\documentclass[sigconf,screen]{acmart}
\acmConference{}
\acmPrice{}
\acmISBN{}
\acmBooktitle{}
\setcopyright{none}
\renewcommand\footnotetextcopyrightpermission[1]{}

\AtBeginDocument{%
  \providecommand\BibTeX{{%
    \normalfont B\kern-0.5em{\scshape i\kern-0.25em b}\kern-0.8em\TeX}}}

\usepackage{adjustbox}
\usepackage{amsmath,amssymb,amsfonts}
\usepackage{algorithmic}
\usepackage{graphicx}
\usepackage{textcomp}
\usepackage{xcolor}
\usepackage{multirow}
\usepackage{mdframed}
\usepackage{tcolorbox}
\usepackage{makecell}
\usepackage{threeparttable}
\usepackage{colortbl}
\usepackage{subfigure}
\usepackage{hhline}
\usepackage{pifont}
\usepackage{xspace}
\usepackage{listings}
\usepackage[frozencache,cachedir=minted-cache]{minted}
\usepackage{hyperref}
\hypersetup{hypertex=true,
            colorlinks=true,
            linkcolor=blue,
            anchorcolor=blue,
            citecolor=blue}
\usepackage{tabularx}
\setminted[]{
  frame=lines,
  framesep=0.5em,
  linenos=true,
  numbersep=0.5em,
  fontsize=\tiny,
  resetmargins=true,
  xleftmargin=1.0em,
  style=friendly,
}
\usepackage{enumitem}

\usepackage{booktabs}

\newcommand{\option}[1]{}
\newcommand{\parabf}[1]{\noindent\textbf{#1}}

\newcommand{\Comment}[1]{}
\newcommand{\CodeIn}[1]{\texttt{#1}}
\newcommand{\yiling}[1]{\textcolor{blue}{Yiling: #1}}

\newcommand{\lingming}[1]{\textcolor{red}{Lingming: #1}}
\newcommand{\civi}[1]{\textcolor{olive}{Ming: #1}}
\newcommand{\jun}[1]{\textcolor{blue}{Jun: #1}}
\newcommand{\yuehan}[1]{\textcolor{orange}{Yuehan: #1}}
\newcommand{\etal}{\textit{et al.}}
\newcommand{\etc}{\textit{etc.}}
\newcommand{\ie}{\textit{i.e.}}
\newcommand{\eg}{\textit{e.g.}}
\newcommand{\overfitprecision}{$Precision$\xspace}
\newcommand{\overfitrecall}{$Recall$\xspace}
\newcommand{\auc}{AUC\xspace}
\newcommand{\prauc}{PR-AUC\xspace}
\newcommand{\avr}{AVR\xspace}
\newcommand{\correctrecall}{$Recall_c$\xspace}
\newcommand{\cpr}{CPR\xspace}
\newcommand{\acc}{ACC\xspace}

\newcommand{\dfj}{Defects4J\xspace}
\newcommand{\dfjold}{Defects4J V1.2\xspace}
\newcommand{\dfjnew}{Defects4J V2.0\xspace}

\newcommand{\tiandata}{$Tian_{v 1.2}$}
\newcommand{\xiongdata}{$Xiong_{v 1.2}$}
\newcommand{\shangwendata}{$Wang_{v 1.2}$}
\newcommand{\dfjoldprapr}{$PraPR_{v 1.2}$}
\newcommand{\dfjnewprapr}{$PraPR_{v 2.0}$}
\newcommand{\dfjmerge}{$Merge_{v 2.0}$}
\newcommand{\dfjbalance}{$Balance_{v 2.0}$}
\newcommand{\dfjdelta}{$PraPR_{v2.0 - v1.2}$}

\newcommand{\apr}{APR\xspace}
\newcommand{\pcc}{PCC\xspace}

\newcommand{\prapr}{PraPR\xspace}
\newcommand{\patchsim}{Patch-Sim\xspace}
\newcommand{\sthree}{S3\xspace}
\newcommand{\ssfix}{ssFix\xspace}
\newcommand{\opad}{Opad\xspace}
\newcommand{\antipattern}{Anti-patterns\xspace}

\newcommand{\simfix}{SimFix\xspace}

\newcommand{\capgen}{CapGen\xspace}

\newcommand{\sumentropy}{Sum Entropy\xspace}
\newcommand{\meanentropy}{Mean Entropy\xspace}
\newcommand{\naturalness}{naturalness-based\xspace}

\definecolor{ggray}{HTML}{eff0f0}
\definecolor{gggray}{HTML}{E8E8E8}
\definecolor{ggggray}{HTML}{BEBEBE}

\newcommand{\tabincell}[2]{\begin{tabular}{@{}#1@{}}#2\end{tabular}}

\usepackage[utf8]{inputenc}
\usepackage[english]{babel}

\usepackage{caption}

\newcounter{finding}
\newcommand{\finding}[1]{\refstepcounter{finding}
 	\vspace{1mm}
	\begin{mdframed}[linecolor=gray,roundcorner=12pt,backgroundcolor=gray!15,linewidth=3pt,innerleftmargin=2pt, leftmargin=0cm,rightmargin=0cm,topline=false,bottomline=false,rightline = false]
		\textbf{Finding \arabic{finding}:} #1
	\end{mdframed}
	\vspace{1mm}
}

\newcommand{\distance}{5pt}
\begin{document}
\date{}

\title{Attention: Not Just Another Dataset for\\ Patch-Correctness Checking}

\author{Jun Yang}
    \affiliation{\institution{University of Illinois at Urbana-Champaign}\country{USA}}
    \email{jy70@illinois.edu}

\author{Yuehan Wang}
    \affiliation{\institution{University of Illinois at Urbana-Champaign}\country{USA}}
    \email{yuehanw2@illinois.edu}
    
\author{Yiling Lou}
    \affiliation{\institution{Fudan University}\country{China}}
    \email{yilinglou@fudan.edu.cn}

\author{Ming Wen}
    \affiliation{\institution{Huazhong University of Science and Technology}\country{China}}
    \email{mwenaa@hust.edu.cn}
    
\author{Lingming Zhang}
    \affiliation{\institution{University of Illinois at Urbana-Champaign}\country{USA}}
    \email{lingming@illinois.edu}

\begin{abstract}
Automated Program Repair (APR) techniques have drawn wide attention from both academia and industry. Meanwhile, one main limitation with the current state-of-the-art APR tools is that patches passing all the original tests are not necessarily the correct ones wanted by developers, i.e., the \emph{plausible patch problem}. To date, various Patch-Correctness Checking (PCC) techniques have been proposed to address this important issue. However, they are only evaluated on very limited datasets as the APR tools used for generating such patches can only explore a small subset of the search space of possible patches, posing serious threats to external validity to existing PCC studies. In this paper, we construct an extensive PCC dataset (the largest manually labeled PCC dataset to our knowledge) to revisit all state-of-the-art PCC techniques. More specifically, our PCC dataset includes 1,988 patches generated from the recent PraPR APR tool, which leverages highly-optimized bytecode-level patch executions and can exhaustively explore all possible plausible patches within its large predefined search space (including well-known fixing patterns from various prior \apr{} tools).\Comment{\lingming{we can further highlight why PraPR is ideal here}} Our extensive study of representative PCC techniques on the new dataset has revealed various surprising findings, including: 1) the assumption made by existing static PCC techniques that correct patches are more similar to buggy code than incorrect plausible patches no longer holds, 2) state-of-the-art learning-based techniques tend to suffer from the dataset overfitting problem, 3) while dynamic techniques overall retain their effectiveness on our new dataset, 
their performance drops substantially on patches with more complicated changes and 4) the very recent naturalness-based techniques can substantially outperform traditional static techniques and could be a promising direction for PCC. Based on our findings, we also provide various guidelines/suggestions for advancing PCC in the near future.

\end{abstract}
\maketitle

\section{Introduction}



Automated Program Repair (\apr)~\cite{DBLP:conf/ssbse/MartinezM18/cardumen, DBLP:conf/icse/DurieuxM16/dynamoth,DBLP:journals/tse/YuanB20/arja, jiang2018shaping} aims to directly fix software
bugs with minimal human intervention and thus can substantially
speed up software development. Recently various \apr
techniques have been proposed and extensively studied in
academia~\cite{DBLP:conf/icse/0001WKKB0WKMT20, DBLP:conf/issta/GhanbariBZ19,DBLP:conf/wcre/LiuK0B19/avatar, DBLP:journals/ese/KoyuncuLBKKMT20/fixminer, DBLP:conf/icse/YiTMBR18, smith2015cure}; furthermore, \apr techniques have also drawn wide
attention from the industry (e.g., Facebook ~\cite{DBLP:conf/icse/MargineanBCH0MM19}, Alibaba~\cite{DBLP:conf/issta/LouGLZZHZ20,benton2020effectiveness},  and
Fujitsu~\cite{DBLP:conf/icse/SahaSP19/hercules,DBLP:conf/kbse/SahaLYP17/elixir}). A typical \apr technique first generates various
candidate patches based on different strategies, such as
template-based~\cite{DBLP:conf/wcre/LiuK0B19/avatar, DBLP:journals/ese/KoyuncuLBKKMT20/fixminer,DBLP:conf/icst/LiuKB0KT19/kpar}, heuristic-based~\cite{DBLP:journals/tse/YuanB20/arja, jiang2018shaping},
constraint-based~\cite{DBLP:conf/icse/XiongWYZH0017/acs, DBLP:conf/ssbse/MartinezM18/cardumen, DBLP:conf/icse/DurieuxM16/dynamoth} and learning-based~\cite{DBLP:conf/issta/LutellierPPLW020/coconut, DBLP:conf/icse/Li0N20/dlfix,chen2019sequencer} ones; then, it validates all the candidate patches via software testing~\cite{jiang2018shaping,DBLP:conf/issta/LutellierPPLW020/coconut,DBLP:conf/issta/GhanbariBZ19},
static analysis~\cite{van2018static}, or even formal verification~\cite{DBLP:conf/kbse/Chen0F17/jaid}. To
date, most \apr systems leverage software testing for patch validation
due to the popularity and effectiveness of testing in practice. \Comment{In this way,
any candidate patches that can pass the original tests are called
\emph{plausible} patches.}


Although test-based patch validation has been shown to be practical
for real-world systems, it suffers from the \emph{test overfitting
issue} -- patches passing all the tests (i.e., \emph{plausible}
patches) may not always be semantically equivalent to the
corresponding developer patches~\cite{qi2015analysis}. The reason is that software
tests can hardly cover all possible program behaviors for real-world
systems. Therefore, it is usually recommended that the developers
should manually inspect the plausible patches to find the final
\emph{correct} ones. However, such manual inspection can be extremely
challenging and time consuming given the large number of potentially
plausible patches and code complexity for real-world
systems~\cite{DBLP:conf/issta/GhanbariBZ19}. To relieve the burden on developers, various techniques
have been proposed to automate Patch-Correctness Checking
(\pcc) -- while static techniques aim to infer patch correctness based
on analyzing the patched code snippets~\cite{le2017s3,DBLP:conf/kbse/XinR17/ssfix,DBLP:conf/icse/WenCWHC18}, dynamic techniques
rely on runtime information \option{collected for the original and patched
code versions }to determine patch correctness~\cite{xiong2018identifying,yang2017better}. In recent
years, researchers have also leveraged the recent advances in deep
code embedding for \pcc~\cite{tian2020evaluating}.


While existing \pcc techniques have shown promising results, they are
mostly studied on limited \pcc datasets since the \apr tools used for generating patches for such datasets can only \emph{explore a very small portion of the possible patch search space} due to the following reasons. First, the leveraged \apr tools have been shown to be overfitting to the studied subjects, e.g., \dfjold ~\cite{DBLP:conf/issta/GhanbariBZ19, RepairThemAll2019}. Such \apr tools usually apply very aggressive patch pruning strategies to fix more bugs on the studied subjects, but may fail to fix any bugs on other subjects, e.g., the recent \simfix/\capgen tools~\cite{DBLP:conf/icse/WenCWHC18,jiang2018shaping} widely adopted in existing \pcc datasets can only fix 0/2 more bugs on a newer version of \dfj with $\sim$200 more bugs~\cite{DBLP:conf/issta/GhanbariBZ19}. In this way, the possible plausible patches that should be generated are also missed by such \apr tools, and thus absent in the resulting \pcc dataset. Second, due to the efficiency issue, most studied \apr tools (e.g., \simfix) terminate patch exploration as soon as they find the first plausible patch. In this way, a large number of possible plausible patches can also be missed by such techniques. 
For example, even including 21 \apr tools, the current largest manually labeled \pcc dataset~\cite{DBLP:conf/kbse/WangWLWQZMJ20}\Comment{Tian \textit{et al.}~\cite{tian2020evaluating} claims their dataset contains up to 1,000 patches, but they include 200+ developer patches which are not expected, so it can not be regarded as the largest.} only has 902 patches for \dfjold in total\Comment{citation for the dataset missing? can we just remove the footnote to save space?}.
Third, most of the \pcc datasets are built on the popular \dfjold, making it unclear whether the findings can generalize to other subjects.  
As a result, it is totally unclear whether the promising
experimental results of existing \pcc techniques can generalize to
more realistic datasets and real-world software development.

\Comment{, e.g., most \pcc studies
leverage the patches generated by ? \apr tools on the \dfj 1.2.0
dataset. There are several limitations for such evaluations. First,
most of the studied \apr tools have been shown to be overfitting to the \dfj
1.2.0 dataset by various independent studies~\cite{prapr,
fse19?}. Therefore, the \pcc results can also be overfitting to the
\dfj 1.2.0 dataset. Second, the studied \apr tools can only explore a
small portion of the possible patch search space, and fail to generate
many plausible patches, e.g., the ? \apr tools together can only
generate ? plausible patches in total for all the 395 available bugs
within \dfj 1.2.0.\lingming{this could be split into two separate reasons: 1) the tools are overfitting and cannot generate more plausible patches, 2) the tools are not efficient and have to stop earlier (e.g., due to time limit or stop at first plausible patch)} Therefore, it is unclear whether the promising
experimental results of prior \pcc techniques can generalize to
other \apr techniques and real-world software development.}


In this paper, we aim to revisit existing state-of-the-art \pcc techniques with a
more extensive and real-world dataset to more faithfully evaluate
their effectiveness in practice. To this end, we first construct an
extensive dataset for \pcc based on the recent \prapr repair
system~\cite{DBLP:conf/issta/GhanbariBZ19}. 
We choose \prapr{} given the following reasons: (1) \emph{its predefined patch search space is large} since it applies
popular well-known fixing templates from various prior \apr work~\cite{pitestpage,papadakis2015metallaxis,schuler2009javalanche}, (2) \emph{it is the only available \apr tool (to our knowledge) that can exhaustively explore the entire predefined patch search space} since it generates and validates patches based on its highly optimized on-the-fly bytecode manipulation.
In this way, \prapr can generate all plausible patches within its clearly defined patch search space, which  can largely avoid the dataset overfitting issue~\cite{DBLP:conf/issta/GhanbariBZ19} and can work as an ideal candidate for constructing \pcc datasets.
We apply \prapr to both the widely studied \dfjold dataset~\cite{DBLP:conf/issta/JustJE14} with 395 bugs and the latest \dfjnew dataset with 11 additional subject systems and 401 additional bugs.
In total, our dataset contains 1,988 plausible patches, including 83 correct patches and 1,905 overfitting patches after manual labeling. 
To our knowledge, this is the largest manually labeled dataset for \pcc. 
\Comment{\yiling{Should we mention the other sub-datasets here?} \jun{I'm not sure but no reviewers challenged this.}\lingming{both are fine with me}}
Then, based on our new datasets, we perform an extensive study
of prior \pcc techniques, including state-of-the-art  static~\cite{le2017s3,DBLP:conf/kbse/XinR17/ssfix,DBLP:conf/icse/WenCWHC18,Tan2016anti, xia2022practical}, dynamic~\cite{yang2017better,xiong2018identifying}, and learning-based~\cite{tian2020evaluating,ye2021ods} \pcc techniques. Our empirical study reveals various surprising findings that the community
should be aware of, and based on the findings, we provide various guidelines/suggestions for further advancing PCC techniques in the near future.
Overall, this paper makes the following contributions:
\Comment{ 1) state-of-the-art embedding-based \pcc
techniques can hardly outperform a random-guessing technique, 2) existing
static \pcc techniques also tend to overfit to the existing patch
datasets, 3) the state-of-the-art dynamic \pcc technique, \patchsim,
can hardly identify any plausible patches from our new
dataset. Furthermore, we also discuss the potential impacts and
practical guidelines that our study provides for future \pcc work.
}
\Comment{1) the assumption made by existing static PCC techniques that correct patches are more similar to buggy code than incorrect plausible patches no longer holds, 2) state-of-the-art learning-based techniques tend to suffer from the dataset overfitting problem and 3) while dynamic techniques overall retain their effectiveness on our new dataset, 
their performance drops substantially when encountering patches with more complicated changes.}

\begin{itemize}[leftmargin=10pt, itemindent=0pt, topsep=0pt]
\item \textbf{Real-world Dataset.} We create a large-scale realistic dataset for \pcc.
To our knowledge, this is the first dataset based on exhaustive patch search space exploration,
as well as the largest manually labeled patch dataset for \pcc to date.
\item \textbf{Extensive Study.} We perform an extensive study of state-of-the-art \pcc techniques,
including static, dynamic, and learning-based ones, on our newly constructed dataset.
In total, our experiments cost 580 CPU days, including static analysis, code embedding, test generation, and test execution for all patches within our dataset.\Comment{ \lingming{put some numbers about the large-scale experiments we did} \jun{1,988 patches generated by \prapr were evaluated by one state-of-art embedding techniques, four static techniques and two dynamic techniques. 44,937 tests were generated to evaluate dynamic tools, with each test running flaky check for five times. @Yuehan, is embedding experiment large or not?}}
\item \textbf{Practical Impacts.} Our empirical study reveals various surprising results
the community should be aware of, as well as providing various practical guidelines for future \pcc work, e.g., 
the dynamic \pcc techniques can no longer work for patches on more complex and real-world bugs
and should be largely improved. 
\Comment{\yiling{replace the link with a citation to save space}}

\end{itemize}

\section{Background and Related Work}
In this section, we first introduce the background on automated program repair and  patch correctness checking, and then motivate our study by systematically revisiting datasets used in the literature.  

\parabf{Automated Program Repair.}~\label{sec:back:apr}
Automated Program Repair (APR)~\cite{ DBLP:conf/ssbse/MartinezM18/cardumen, DBLP:conf/icse/DurieuxM16/dynamoth,DBLP:journals/tse/YuanB20/arja, jiang2018shaping,DBLP:conf/wcre/LiuK0B19/avatar, DBLP:journals/ese/KoyuncuLBKKMT20/fixminer, smith2015cure, DBLP:conf/icse/YiTMBR18} aims at fixing bugs with minimal human intervention\Comment{, which generates patches for the buggy program and outputs the ones satisfying desired behaviors}. Due to its promising future, researchers have proposed various APR techniques, which can be categorized according to how patches are generated: (1) \textit{Heuristic-based APR}~\cite{le2012genprog,qi2014strength,yuan2018arja} leverages heuristics \option{(e.g., random searching~\cite{qi2014strength} or genetic programming~\cite{le2012genprog})} to explore the search space of patches; (2) \textit{Template-based APR}~\cite{DBLP:conf/wcre/LiuK0B19/avatar, DBLP:journals/ese/KoyuncuLBKKMT20/fixminer,DBLP:conf/icst/LiuKB0KT19/kpar} incorporates patterns summarized from historical developer patches to guide patch generation; (3) \textit{Constraint-based APR}~\cite{DBLP:conf/icse/XiongWYZH0017/acs, DBLP:conf/ssbse/MartinezM18/cardumen, DBLP:conf/icse/DurieuxM16/dynamoth} leverages \option{symbolic execution and }constraint solving to directly synthesize patches\Comment{ conditional or assignment expressions}; (4) \textit{Learning-based APR}~\cite{DBLP:conf/issta/LutellierPPLW020/coconut, DBLP:conf/icse/Li0N20/dlfix,chen2019sequencer} utilizes learning techniques (e.g., neural machine translation~\cite{bahdanau2014neural}\option{ and Large Language Model~\cite{xia2022less})} to generate patches. APR tools then leverage software testing~\cite{jiang2018shaping,DBLP:conf/issta/LutellierPPLW020/coconut,DBLP:conf/issta/GhanbariBZ19},
static analysis~\cite{van2018static}, or even formal verification~\cite{DBLP:conf/kbse/Chen0F17/jaid} to validate the patches.

\parabf{Patch Correctness Checking.}~\label{sec:back:pcc}
Most APR techniques use software tests for patch validation, regarding the patches that can pass all tests as potentially correct. However, such an assumption can be problematic, since tests often fail to detect all possible bugs\Comment{with limited fault-detection capability in practice}. Therefore, a plausible patch may be overfitting to the existing tests and yet be incorrect. To address such test-overfitting issue, various Patch Correctness Checking (\pcc{})~\cite{DBLP:conf/icse/LeB00LP19} techniques have been proposed to differentiate \textit{overfitting patches} from \textit{correct patches}. 

There are two application scenarios for \pcc{} techniques, with \textit{oracle patch information} or without. In the \textit{oracle-based} scenario, the plausible patches that exhibit differently from the oracle patch are deemed as overfitting. 
\Comment{For example, existing oracle-based \pcc{} techniques often generate new tests (e.g., Evosuite~\cite{Fraser2011EvoSuite} or Randoop~\cite{pacheco2007randoop}) or invariants (e.g., Daikon~\cite{ErnstCGN2001:daikon}) to characterize run-time behaviors of the oracle and plausible patches.}\Comment{ Oracle-based \pcc{} techniques can evaluate the correctness of generated plausible patches when the oracle patch is given, thus mainly serving as effectiveness assessment for APR techniques.} Note that \textit{oracle-based} \pcc{} is usually used for \apr experimentation, and cannot be applied to real-world bug fixing where the oracle patch information is unavailable. In contrast, \textit{oracle-absent} \pcc{} can identify potential overfitting patches without oracle, thus can be applied in real-world bug fixing\Comment{which can be incorporated into patch generation and enable more powerful APR}. Therefore, we only focus on the \textit{oracle-absent} techniques.

According to whether dynamic patch execution information is required, traditional \pcc{} techniques can be categorized into \textit{static} and \textit{dynamic} techniques. In  particular, static techniques leverage static code features~\cite{DBLP:conf/kbse/XinR17/ssfix, le2017s3, DBLP:conf/icse/WenCWHC18}, pre-defined anti-patterns~\cite{Tan2016anti} or even code naturalness~\cite{xia2022practical} to predict patch correctness. Dynamic techniques leverage run-time information, such as crash or memory-safety issues used in \opad{}~\cite{yang2017better} and test execution traces used in \patchsim{}~\cite{xiong2018identifying}. 
Recently, researchers have also started exploring advanced machine learning and deep learning techniques for \pcc. Seminal research~\cite{ye2021ods} characterizes patch code with pre-defined features, while more recent work~\cite{tian2020evaluating,codeembedding} directly encodes patch code via embedding models (e.g., BERT~\cite{DBLP:journals/corr/abs-1810-04805}) and learns to predict the probability of each patches being overfitting.

\begin{table*}[htb]
	\centering
	\small
	\renewcommand{\arraystretch}{1.0}
    \setlength{\tabcolsep}{3pt}
	\caption{Datasets used in existing \pcc{} work}\label{table:revisit}
	\begin{adjustbox}{width=2.0\columnwidth}
	\begin{tabular}{c|c|c|c|c|c|c}
	\toprule
	\multirow{2}{*}{\textbf{Work}} & 	\multirow{2}{*}{\textbf{Studied technique}} & 	\multirow{2}{*}{\textbf{Scale}} & 	\multirow{2}{*}{\textbf{Ratio}} & 	\multicolumn{3}{c}{\textbf{Dataset source}}  \\ \cline{5-7}
	
	&&& (correct/overfitting)& \textbf{Subject source} & \textbf{Patch source} & 	\textbf{Early stop}\\ \hline
	
Tan~\etal{}~\cite{Tan2016anti} & \tabincell{c}{\antipattern{}} & 289 & 30/259 & CoREBench~\cite{bohme2014corebench}, GenProg~\cite{le2012genprog} &
\tabincell{c}{\textbf{ All from APR tools} \\ GenProg, mGenProg~\cite{Tan2016anti}, SPR~\cite{long2015staged}, mSPR~\cite{Tan2016anti}} & 4/4 \\ \hline

    Xin~\etal{}~\cite{10.1145/3092703.3092718} & \tabincell{c}{ DiffTGen } & 89 & 10/79 & \tabincell{c}{\dfjold{}} & \tabincell{c}{\textbf{All from APR tools} \\ jGenProg, jKali, NPol and HDRepair} & 4/4 \\ \hline
	
	Le~\etal{}~\cite{le2017s3} & \tabincell{c}{\sthree{}} & 85 & 25/60 & \tabincell{c}{100 bugs from 62 subjects} \Comment{100 real world bugs}  &
\tabincell{c}{\textbf{All from APR tools} \\ \sthree{}, Enumerative~\cite{reynolds2015counterexample}, CVC4~\cite{reynolds2015counterexample}, Angelix~\cite{mechtaev2016angelix}} & 4/4 \\ \hline
	
	Yang~\etal{}~\cite{yang2017better} & \tabincell{c}{\opad{}} & 449 & 22/427 & \tabincell{c}{45 bugs from 7 subjects}  &
\tabincell{c}{\textbf{All from APR tools} \\ GenProg, AE~\cite{weimer2013leveraging}, Kali~\cite{qi2015analysis}, and SPR} & 4/4 \\ \hline
	
	Xin~\etal{}~\cite{DBLP:conf/kbse/XinR17/ssfix} & \tabincell{c}{ \ssfix{}} & 153 & 122/31 & \tabincell{c}{\dfjold{} \\ \textit{exclude Mockito}}  &
	\tabincell{c}{\textbf{All from APR tools} \\ \ssfix{}, jGenProg, jKali~\cite{martinez2016astor}, Nopol, HDRepair~\cite{le2016history}, ACS} & 5/5 \\ \hline
	
	Xiong~\etal{}~\cite{xiong2018identifying} & \tabincell{c}{ \patchsim{}} & 139 & 29/110 & \tabincell{c}{\dfjold{} \\ \textit{exclude Closure, Mockito}}  &
	\tabincell{c}{\textbf{All from APR tools} \\ jGenPro~\cite{martinez2016astor}, Nopol~\cite{xuan2017nopol}, jKali, ACS~\cite{DBLP:conf/icse/XiongWYZH0017/acs}, HDRepair} & 5/5 \\ \hline
	
	Ye~\etal{}~\cite{ye2021automated} & \tabincell{c}{ RGT} & 638 & 257/381 & \tabincell{c}{\dfjold{}} & \tabincell{c}{\textbf{All from APR tools} \\ ACS, Arja, CapGen, etc., 14 tools in total } & 10/14 \\ \hline
	
	Wen~\etal{}~\cite{DBLP:conf/icse/WenCWHC18} & \tabincell{c}{ CapGen~\cite{DBLP:conf/icse/WenCWHC18}} & 202 & 28/174 & \tabincell{c}{\dfjold{} \\ \textit{exclude Closure, Mockito}}  &
	\tabincell{c}{\textbf{All from APR tools} \\ CapGen} & 0/1 \\ \hline
	
	Xin~\etal{}~\cite{10.1109/GI.2019.00012} & \tabincell{c}{ sharpFix~\cite{xin2019revisiting}} & 82 & 56/26 & \tabincell{c}{Bugs.jar \\ \textit{127 bugs from 7 projects in Bugs.jar}}  &
	\tabincell{c}{\textbf{All from APR tools} \\ sharpFix, \ssfix{}~\cite{DBLP:conf/kbse/XinR17/ssfix}} & 2/2 \\ \hline
	
	Tian~\etal{}~\cite{tian2020evaluating} & \tabincell{c}{Embedding-based}& 1,000 & 468/532 & \dfjold{} &
	\tabincell{c}{\textbf{778 from APR tools} \\ RSRepair-A~\cite{yuan2018arja}, jKali, ACS, SimFix~\cite{jiang2018shaping}, TBar~\cite{liu2019tbar}, etc., 17 tools in total \\ \textbf{232 from developer patches}  \textit{(*Supplemented to balance the dataset)}} & 12/17 \\ \hline
	

	Wang~\etal{}~\cite{DBLP:conf/kbse/WangWLWQZMJ20} & \tabincell{c}{
	12 \pcc{} techniques}& 902 & 248/654 & \dfjold{} &
	\tabincell{c}{\textbf{All from APR tools} \\ jGenProg, DynaMoth~\cite{durieux2016dynamoth}, SequencR~\cite{chen2019sequencer}, etc., 21 tools in total } & 16/21 \\ \hline
	
	Lin~\etal{}~\cite{lin2022context} & \tabincell{c}{Cache} & 49,694 & 25,589/24,105 & RepairThemAll~\cite{RepairThemAll2019}, ManySStuBs4J~\cite{karampatsis2020often} & \tabincell{c}{\textbf{Overfitting patches from \apr tools and correct patches from developer patches} \\ jGenProg, jKali, Nopol, etc., 11 \apr tools in total and human patches} & 11/11 \\ \hline
	
	Ye~\etal{}~\cite{ye2021ods} & \tabincell{c}{ODS} & 10,302 & 2,003/8,299 & RepairThemAll, \dfjnew developer patches & \tabincell{c}{\textbf{Overfitting patches from \apr tools and correct patches from developer patches} \\ jGenProg, jKali, Nopol, etc., 11 \apr tools in total and human patches} & 11/11 \\

	\bottomrule
	\end{tabular}
	\end{adjustbox}
\end{table*}

\parabf{Existing \pcc{} Datasets.}~\label{sec:back:dataset}
We revisit the existing peer-refereed literature on automated program repair and \pcc{} based on the APR review~\cite{livingapr} before 2022, and summarize the datasets used in their evaluation. Table~\ref{table:revisit} presents the characteristics of datasets used in existing work. \Comment{In particular, Column ``Studied technique'' presents the \pcc{} techniques proposed/studied in each work. Column ``Scale'' presents the total number of patches; Column ``Ratio'' presents the ratio between correct patches and overfitting patches in each dataset. We further show how patches are collected in Column ``Dataset source'', including the subjects, the number of developer patches or APR-generated patches.} Note that for the patches generated by APR tools, we check whether the early stop mechanism is enabled during APR procedure\Comment{, i.e., for each bug, the \apr tools will stop as soon as the first plausible patch is returned\Comment{only the first plausible patch would be reported and included into the dataset}. The column ``Early stop'' presents the number of APR tools with early stop mechanism.} as shown in Column ``Early stop''. From the table, we can find that although involving various APR tools, subjects, and patches, existing \pcc{} datasets can still be insufficient for evaluating \pcc, since they can only explore a very small portion of the possible patch search space due to the following reasons.

\textit{First, existing datasets mainly include the plausible patches generated by overfitting APR tools.}  As shown in recent work~\cite{RepairThemAll2019, DBLP:conf/issta/GhanbariBZ19}, most existing APR tools are overfitting to the  \dfjold{}~\cite{DBLP:conf/issta/JustJE14} benchmark. Such APR tools often leverage aggressive and specific patch pruning strategies so as to fix more bugs, but may suffer from the dataset-overfitting issue~\cite{smith2015cure}\Comment{which cannot be general to other subjects and thus may fail to fix any new bugs}. For example, the recent \simfix/\capgen tools~\cite{DBLP:conf/icse/WenCWHC18,jiang2018shaping} can only fix 0/2 more bugs on additional $\sim$200 more bugs~\cite{DBLP:conf/issta/GhanbariBZ19}; also, 11 APR tools have been shown to perform substantially worse on other benchmarks than the widely-used \dfjold{}~\cite{RepairThemAll2019}. In this way, the possible plausible patches that should be generated are missed, and thus absent in the resulting \pcc dataset.

\textit{Second, existing datasets only include the first one/few plausible patches generated by APR tools due to efficiency issue.} \apr tools (e.g., \simfix~\cite{jiang2018shaping}) enable \textit{early stop} mechanism, terminating patch execution as soon as they find the first plausible patch for the sake of efficiency. As shown in the table, most APR tools in existing \pcc{} datasets enable \textit{early stop}. In this way, existing \pcc{} dataset may miss a large number of possible plausible patches that can be potentially generated by these APR tools. 
According to Noller~\etal{}'s study~\cite{noller2022trust}, developers expect \apr quickly(30-60min) generate 5-10 patches each bug. But among the 16 \apr tools investigated by Liu~\etal{}~\cite{DBLP:conf/icse/0001WKKB0WKMT20}, the most effective one, \simfix, set a timeout of 300 min for each bug but only generated 68 patches on \dfjold. Therefore, \apr tools for previous datasets are neither practical nor efficient. In fact, even including 21 \apr tools, the existing largest manually labeled \pcc dataset~\cite{DBLP:conf/kbse/WangWLWQZMJ20} has only 902 plausible patches for \dfjold; meanwhile a single APR (i.e., SequenceR~\cite{chen2019sequencer}) can only generate at most 73 patches. 

\textit{Third, most \pcc datasets are built on the popular \dfjold (old version), making it unclear whether the findings can generalize to other subjects.}
As shown in the table, the majority of existing \pcc{} datasets are generated from the subjects in the most widely used benchmark \dfjold{}. Therefore, it remains unknown how existing \pcc{} techniques perform on patches from other subjects. 

\Comment{\yiling{we do not need so much details here. Just mentioned they are not manually labeled and there are some threats. Two sentences are fine.}}

\Comment{\yuehan{We may shorten this part... Actually I think this issue is kind of overlapped with the first issue we mentioned before (the dataset includes the plausible patches), considering they collected these so-called big datasets based on overfitting APR tools.}}


To address the limitations above, in this work, we construct a more extensive and real-world dataset so as to revisit all representative \pcc techniques. In particular, our dataset is built on the plausible patches generated by the recent byte-code level APR tool \prapr{}~\cite{DBLP:conf/issta/GhanbariBZ19}. We choose \prapr since 1) its predefined patch search space is large since it applies popular well-known fixing templates from various prior \apr work~\cite{pitestpage,papadakis2015metallaxis,schuler2009javalanche}, and 
2) it is the baseline \apr tool that can exhaustively explore the predefined patch search space due to its highly
optimized on-the-fly bytecode manipulation.
In this way, \prapr can generate all plausible patches within the clearly defined patch search space, which other APR tools fail to reach. Furthermore, in addition to the most widely used \dfjold dataset~\cite{DBLP:conf/issta/JustJE14}, we further
apply \prapr to the most recent version of \dfj dataset, i.e., \dfjnew  with 11 additional subject systems and 401 additional bugs. 
\Comment{We can mention "to avoid the bias which can be brought by entirely focusing on a single tool, we also consider some sub-datasets (see in Section 3.3)" to emphasize that we have a comprehensive thinking before the reviewers may challenge us due to the single application of PraPR here. }

\section{Dataset Construction}

\Comment{\yiling{I rewrite the whole section, plz double check.}\lingming{made a quick pass, looks great!}}
\subsection{Subjects}
Our datasets are constructed with patches generated for the subjects in \dfj{}~\cite{DBLP:conf/issta/JustJE14} due to the following reasons: 1) \dfj contains hundreds of real bugs for real-world projects and has become the most widely studied \apr benchmark in the literature, 2) most prior \pcc{} studies were performed on \dfj~\cite{DBLP:conf/issta/JustJE14,DBLP:conf/kbse/WangWLWQZMJ20,10.1145/3092703.3092718,xiong2018identifying,ye2021automated,tian2020evaluating},
\Comment{\civi{may add some citations here.}\jun{added}} thus enabling a more direct/fair comparison with prior work. We include all subjects from \dfjold{} (with 395 bugs from 6 projects) since it is the most widely used \dfj{} version in prior \apr and \pcc work~\cite{DBLP:conf/issta/JustJE14,DBLP:conf/kbse/WangWLWQZMJ20,10.1145/3092703.3092718,xiong2018identifying,ye2021automated,DBLP:conf/kbse/XinR17/ssfix,DBLP:conf/icse/WenCWHC18,tian2020evaluating}.\Comment{\civi{add some citation} \jun{done}} Furthermore, we also study the latest \dfjnew{}, which includes 11 additional projects with 401 additional bugs, to study the dataset overfitting issue of existing \pcc work, i.e., whether the prior \pcc experimental results can generalize to the newer \dfj version. \Comment{Table~\ref{tab:d4j_programs} presents the basic statistics of both \dfj versions.} Please refer to our website~\cite{replication} for detailed statistics of \dfj versions.

\Comment{To date, there are two versions of \dfj{}: \dfjold{} with 395 bugs and \dfjnew{} with additional 401 bugs. To the best of our knowledge, all existing \pcc{} work is focusing on patches generated from \dfjold{} \jun{maybe some citation here?}, while it remains unknown how \pcc{} performs on additional bugs in \dfjnew{},
Therefore, our study considers patches generated from both versions, so that we can not only compare our findings with previous work on the same subject benchmark, but also explore the effectiveness of \pcc{} techniques on more additional bugs. Table~\ref{tab:d4j_programs} presents the basic statistics of both subject benchmarks. }

\subsection{Patch Collection}    
\parabf{Plausible Patch Collection.} For each studied buggy version from \dfj, we first run \prapr on it to generate all possible plausible patches (i.e., the patches passing all the tests). 
Since \prapr{} is based on bytecode manipulation, the resulting plausible patches are in the bytecode level. We further decompile these bytecode-level plausible patches into source-code-level patches since existing PCC techniques work at the source-code level.    
We make the following efforts to ensure the decompilation process as precise as possible.
First, we configure \prapr{} to include all the required debugging information in the resulting bytecode-level patches. Second, we leverage the state-of-the-art decompiler JD-Core~\cite{JDCore} to decompile patched bytecode files\Comment{(with the corresponding line number information attached)}. Third, we only locate the patched line in the decompiled file (note that \prapr{} only changes one line on bytecode level patches), and patch this line into the original buggy source file with the help of debugging information to obtain a potential patch at the source-code level.
Finally, we perform the sanity check to ensure the decompiled source-code-level patches indeed pass all the tests. Finally, we obtain 1,988 plausible patches.

\parabf{Patch Correctness Identification.} 
For each plausible patch, we then manually determine its correctness by comparing it to the developer patch. In particular, we follow the labeling procedure in previous work~\cite{DBLP:conf/icse/0001WKKB0WKMT20} and the patches that satisfy the following criteria are labeled as correct: (1) the patches that are syntactically identical to the developer patches, or (2) the patches that are semantically equivalent to the developer patches according to the rules summarized from existing work\Comment{The rules are summarized from the correct patches generated by multiple existing APR tools More details are in~\cite{DBLP:conf/icse/0001WKKB0WKMT20}.}~\cite{DBLP:conf/icse/0001WKKB0WKMT20}. Otherwise, the patches are labeled as overfitting. 
We involve multiple participants with 3+ years \Comment{\civi{over how many years?}} Java development experience in the manual labelling procedure: two participants first label each plausible patch individually, and a third participant would be introduced to resolve the conflicts by discussing based on the corresponding bug report or issue link. In total, among all the 1,988 new plausible patches, 83 are correct and 1,905 are overfitting ones.

\Comment{\begin{table}[t!]
      \setlength\tabcolsep{8.1pt} 
      \def\arraystretch{1.0}
      \scriptsize
      \caption{Subject benchmarks}
      \centering
      \begin{tabular}{l|l|r|r|r}
    \toprule
    Identifier             & Project Name              & Bugs & Tests & LoC   \\
    \midrule
    \textit{Chart}         & JFreeChart                & 26   & 2,193 & 327K   \\
    \textit{Time}          & Joda-Time                 & 27   & 4,130 & 142K   \\
    \textit{Mockito}       & Mockito framework         & 38   & 1,379 & 94K   \\
    \textit{Lang}          & Apache commons-lang       & 65   & 2,291 & 111K   \\
    \textit{Math}          & Apache commons-math       & 106  & 4,378 & 302K   \\
    \textit{Closure}       & Google Closure compiler   & 133  & 7,911 & 379K   \\ \hline
   \multicolumn{5}{c}{\textbf{\dfjold}}  \\
    \midrule
    \textit{Cli}           & Commons-cli   & 39   &94   & 8K    \\ 
    \textit{Codec}           & Commons-codec   & 18   & 206   & 12K    \\ 
    \textit{Collections}    & Commons-collections & 4   & 15,393   & 113K    \\ 
    \textit{Compress}           & Commons-compress   & 47   & 73   & 16K    \\ 
    \textit{Csv}           & Commons-csv  & 16   & 54   & 4K    \\ 
    \textit{Gson}           & Gson   & 18   & 720   & 30K    \\ 
    \textit{JacksonDatabind}           & Jackson-databind   & 112   & 1,098   & 111k    \\ 
    \textit{JacksonCore}           & Jackson-core  & 26   & 206   & 36k    \\ 
    \textit{JacksonXml}           & Jackson-dataformat-xml  & 6   & 138   & 14K    \\ 
    \textit{Jsoup}           & Jsoup  & 93   & 139   & 6K    \\ 
    \textit{JXPath}           & Apached commons-jxpath   & 22   & 308   & 36K    \\ \hline
    \multicolumn{5}{c}{\textbf{\dfjnew}}  \\
    \bottomrule
\end{tabular}
\label{tab:d4j_programs}
      \label{tab:d4j_programs}
\end{table}}

\Comment{
\begin{table}[t!]
      \setlength\tabcolsep{8.1pt} 
      \scriptsize
      \caption{Mapping from \prapr mutators to TBar fix patterns and tools implementing the pattern}
      \centering
\begin{tabularx}{9cm}{l|X}
    \toprule
    Prapr Mutator          & Implemented by APR Tools               \\
    \hline
    Argument Propagation    & PAR, HDRepair, ssFix, ELIXIR, FixMiner, SOFix, SketchFix, CapGen, and SimFix                 \\
    \hline
    Return Value            & ELIXIR, SketchFix, and HDRepair \\
    \hline
    Constructor Call        & AVATAR \\
    \hline
    Inline Constants        & HDRepair, S3, FixMiner, SketchFix, CapGen, SimFix and ssFix \\
    \hline
    Member Variable         & S3, SOFix, FixMiner, SketchFix, CapGen, SimFix, AVATAR, and ssFix \\
    \hline
    Switch                  & S3, SOFix, FixMiner, SketchFix, CapGen, SimFix, AVATAR, and ssFix \\
    \hline
    Method Call             & PAR, HDRepair, ssFix, ELIXIR, FixMiner, SOFix, SketchFix, CapGen, and SimFix   \\
    \hline
    Invert Negatives        & HDRepair, S3, FixMiner, SketchFix, CapGen, SimFix and ssFix \\
    \hline
    Arithmetic Operator     & HDRepair, ssFix, ELIXIR, S3, jMutRepair, SOFix, FixMiner, SketchFix, CapGen, SimFix, AVATAR, and PAR \\
    \hline
    Conditional             & PAR, ssFix, S3, HDRepair, ELIXIR, SketchFix, CapGen, SimFix, and AVATAR \\
    \hline
    Variable Replacement    & S3, SOFix, FixMiner, SketchFix, CapGen, SimFix, AVATAR, and ssFix \\ 
    \hline
    Field Replacement       & S3, SOFix, FixMiner, SketchFix, CapGen, SimFix, AVATAR, and ssFix \\
    \hline
    Method Replacement      & PAR, HDRepair, ssFix, ELIXIR, FixMiner, SOFix, SketchFix, CapGen, and SimFix \\
    \hline
    Type Replacement        & PAR, ELIXIR, FixMiner, SOFix, CapGen, SimFix, AVATAR, and HDRepair \\
    \hline
    Field Guard             & PAR, ELIXIR, NPEfix, Genesis, FixMiner, AVATAR, HDRepair, SOFix, SketchFix, CapGen, and SimFix \\
    \hline
    Method Guard            & PAR, ELIXIR, NPEfix, Genesis, FixMiner, AVATAR, HDRepair, SOFix, SketchFix, CapGen, and SimFix \\
    \hline
    Pre/Post-Condition      & PAR, ELIXIR, NPEfix, Genesis, FixMiner, AVATAR, HDRepair, SOFix, SketchFix, CapGen, and SimFix \\
    \bottomrule
\end{tabularx}
\label{tab:mut_mapping}
      \label{tab:mut_mapping}
\end{table}
}

\begin{table}[t!]
	\centering
	\small
	\renewcommand{\arraystretch}{1.1}
	\caption{Our patch datasets}\label{table:patchdataset}
	\begin{adjustbox}{width=1.0\columnwidth}
	\begin{tabular}{l|l|l|r|r|r}
	\hline
    \textbf{Dataset ID} & \textbf{Subjects} & \textbf{APR Tools} & \textbf{\#Patch} & \textbf{\#Overfit} & \textbf{\#Correct} \\ \hline
        
    \dfjoldprapr{} &  \dfjold{} &\prapr{} & 1,311 &  1,264 & 47  \\ 
    \dfjnewprapr{} &  \Comment{\dfjold{} +} \dfjnew{}&\prapr{}  & 1,988 & 1,905 & 83 \\
    \dfjmerge{} & \dfjnew{}& \prapr{} + 21 tools & 2,760 & 2,489 & 271 \\
    \dfjbalance{} & \dfjnew{}& \prapr{} + 21 tools & 542 & 271 & 271 \\
	\hline
	\end{tabular}
	\end{adjustbox}
\end{table}

\subsection{New Patch Datasets}
The plausible patches above constitute our main dataset \dfjnewprapr{}, with 1,988 plausible patches generated by \prapr{} on \dfjnew{}. Based on the main dataset, we further construct three sub-datasets for more thorough study of PCC work. Table~\ref{table:patchdataset} presents the details.

\Comment{We then construct four datasets with patches generated by \prapr{} and other APR tools on both versions of \dfj{}. Table~\ref{table:patchdataset} presents the statistics of four datasets. We then introduce the construction procedure in detail. 
\civi{Is there a better way to denote the symbol of datasets? It reads unclear especially in Tables. They all contain ``Defects4J'', which can be omitted. Denoting it as $PraPR^{V1.2}$ is fine with me.}}

\textbf{\dfjoldprapr{}}.
\Comment{the description about 2.0 is still not consistent across the paper: lets make it clear that 2.0 include bugs from 1.2. If you only want additional bugs from 2.0, use 2.0-1.2}
This dataset includes all 1,311 plausible patches generated by \prapr{} on \dfjold{}, separated from the main dataset to compare with prior \pcc{} studies on the same \dfj version.
\Comment{ \lingming{we should somehow highlight our (manual) efforts in building the dataset}\jun{Add some details in last paragraph.}}


\textbf{\dfjmerge{}}. Though \prapr contains overlapping patch fixing patterns with many other \apr tools, 
it is still important to consider patches from other \apr tools for more comprehensive \pcc evaluation. 
Hence, we further combine our \dfjnewprapr{} with the existing largest labeled dataset derived from \apr tools, i.e., \shangwendata{}~\cite{DBLP:conf/kbse/WangWLWQZMJ20} that includes 902 plausible patches generated by 21 other \apr tools. 
After carefully removing the duplicates between \shangwendata{} and our dataset, we finally obtain 2,763 plausible patches in this merged dataset. 

\textbf{\dfjbalance{}}. Table~\ref{table:patchdataset} shows that the \dfjmerge{} dataset is largely imbalanced with mostly overfitting patches.
While \pcc studies frequently leverage imbalanced datasets\option{ (since current \apr tools cannot generate correct patches for many bugs)}, we further construct a balanced dataset based on 
\dfjmerge{} for more thorough evaluations. 
Specifically, we keep all correct patches from \dfjmerge{} and randomly sample the same number of overfitting patches. \Comment{To avoid including an overwhelming portion of the overfitting patches from the previous dataset (i.e., \shangwendata{}),}
To mitigate the bias in randomness, we repeat the random sampling process for ten times and present the average results on \dfjbalance{}.

To our knowledge, the datasets newly constructed in this work are the largest manually labeled datasets for evaluating \pcc{}. For example, \prapr{} alone generates 1,988 plausible patches while the existing largest \pcc dataset has only 902 plausible patches (Lin~\etal~\cite{lin2022context} and Ye~\etal~\cite{ye2021ods} have datasets with \textasciitilde50,000 and \textasciitilde10,000 patches but they are not manually labeled). In addition, merging the patches generated by \prapr with the existing datasets could further form a larger labeled dataset (i.e., \dfjmerge{}). The large scale of our new datasets is credited to the high efficiency and exhaustive search strategies of \prapr{}, which help collect many plausible patches that are missed by the other \apr{} tools due to their efficiency issues or their early termination\option{ after finding the first plausible patch}.
For example, \emph{\prapr{} alone generates 39 unique plausible patches for the Defects4J bug Math-28, while the other existing 21 \apr{} tools generate 7 plausible patches in total (5 of them are unique)}. The main reason could be that other tools early stop after finding the first plausible patch while \prapr exhaustively explores the whole search space. Therefore, at least 34 out of the 39 plausible patches generated by \prapr{} are not visited by the existing 21 \apr{} tools. 

\section{Study Design}
\subsection{Research Questions}

\Comment{In this study, we aim to answer the following research questions.}
Based on our newly constructed datasets, we revisit all state-of-the-art \pcc{} techniques via the following research questions:

\begin{itemize}[leftmargin=10pt, itemindent=0pt, topsep=0pt]
    \item \textbf{RQ1:} How does \emph{static} \pcc{} perform on our  datasets?
    \item \textbf{RQ2:} How does \emph{learning-based} \pcc{} perform on our datasets?
    \item \textbf{RQ3:} How does \emph{dynamic} \pcc{} perform on our datasets?

\end{itemize}

\subsection{Studied Techniques}~\label{sec:setup:tech}
Our study selects existing state-of-the-art techniques that are designed for or can be adapted to the \pcc{} task.
Specifically, the selected techniques can be broadly categorized into three categories, including \textit{static}, \textit{dynamic}, and \textit{learning-based} techniques.
We only include those techniques that do not require the oracle information (i.e., the developer patches) since the practical usefulness of those techniques requiring the oracle information is compromised~\cite{DBLP:conf/kbse/WangWLWQZMJ20}. 


\Comment{\lingming{we should mention somewhere we only target techs not requiring oracle info!}}

\subsubsection{Static Techniques}~\label{sec:setup:tech:static}
Wang \etal~have empirically investigated the effectiveness of static features extracted from three tools~\cite{DBLP:conf/kbse/WangWLWQZMJ20}, namely ssFix~\cite{DBLP:conf/kbse/XinR17/ssfix}, S3~\cite{le2017s3} and CapGen~\cite{DBLP:conf/icse/WenCWHC18}, and then utilized such features to check patch correctness.
We use identical experiment settings and methodology to assess patch correctness in new dataset.

\textbf{S3:} S3 proposed six features to measure the syntactic and semantic distance between patched code and the original buggy code~\cite{DBLP:conf/sigsoft/LeCLGV17/s3}.
Among them, \textit{AST differencing} and  \textit{cosine similarity}\Comment{inconsistent capitalization among three tools! \lingming{if s3 hsa similarity, why higher scores will be deemed as overfitting?}, and \textit{locality of variables and constants}\jun{we reused statements from Shangwen's paper, here this similarity is actually equivalent with distance, i.e., smaller distance means larger similarity}} are utilized to prioritize and identify correct patches.
Specifically, the sum of these features is computed\option{ and further used} as the suspiciousness score.
Patches with higher scores are more likely to be overfitting according to the previous study~\cite{8918958} which claims correct patches are more similar to the original buggy code, and thus possess fewer modifications.
In addition, we exclude the other three features in this study following prior work~\cite{DBLP:conf/kbse/WangWLWQZMJ20}\Comment{\lingming{did \cite{DBLP:conf/kbse/WangWLWQZMJ20} also exclude these? if so add ``following prior work~\cite{DBLP:conf/kbse/WangWLWQZMJ20}''}\jun{added}}.
Specifically, \textit{model counting} and \textit{output coverage} are excluded since they cannot be generalized to all the generated patches\Comment{, as pointed out by a prior study}~\cite{DBLP:conf/kbse/WangWLWQZMJ20}.
Besides, \textit{Anti-patterns} utilized in S3 is excluded here since we utilize it as a stand-alone tool, following the previous study~\cite{DBLP:conf/kbse/WangWLWQZMJ20}.

\textbf{ssFix:} ssFix focused on the token-based syntax representation of code to identify syntax-related code fragments to generate correct patches.
Specifically, \textit{structural token similarity} and \textit{conceptual token similarity} are calculated to measure the similarity between the buggy code and patched code.
Similar to \textit{S3}, the sum of these features is used to rank patches.
Patches with higher scores are ranked higher.

\textbf{CapGen:} CapGen designed three context-aware features to prioritize correct patches over overfitting ones, namely the \textit{genealogy similarity}, \textit{variable similarity}, and \textit{dependency similarity} respectively.
Although such features are not initially proposed for \pcc, they can still be adapted to assess patch correctness from the view of static features.
Specifically, following the existing study~\cite{DBLP:conf/kbse/WangWLWQZMJ20}, the product of these similarity scores is used to rank and prioritize patches, and patches with higher scores are considered more likely to be correct.

For the above three static techniques, we apply the Top-N strategy to identify correct patches following prior work~\cite{DBLP:conf/kbse/WangWLWQZMJ20}. 
Specifically, the Top-N prioritized patches are labeled as correct while others as overfitting, where N is the number of correct patches in our dataset following prior studies~\cite{DBLP:conf/kbse/WangWLWQZMJ20}. 
To mitigate the effect of N, we also show the average ranking of correct patches per bug (i.e., denoted as \avr) to evaluate the ability of prioritizing correct patches (detailed in Section~\ref{sec:metric}).

\textbf{\antipattern{}:} \antipattern{} provides a set of generic forbidden transformations, which can be applied on search-based repair tools~\cite{Tan2016anti}. 
It has been shown that \textit{\antipattern{}} could help obtain program patches with higher qualities with minimal effort. 
To apply these \antipattern{} on our own dataset, we first map the \antipattern{} to \prapr's mutation rules~\cite{DBLP:conf/issta/GhanbariBZ19} through manual inspection.
If the rules fall into a specific pattern, we deem those patches generated by the mutation rule as overfitting.
Such a strategy is the same as that adopted by the existing study~\cite{DBLP:conf/kbse/WangWLWQZMJ20}. 
In this way, we observe that patches generated by \prapr mainly fall into four of the seven anti-patterns, including A1: \textit{Anti-delete CFG Exit Node}, A2: \textit{Anti-delete Control Statement}, A5: \textit{Anti-delete Loop-Counter} and A6: \textit{Anti-append Early Exit}. 
Therefore, we classify the patches that fall into these rules as overfitting and the others as correct.
\Comment{\textbf{Code naturalness:}}

\textbf{Code naturalness:} Cross entropy~\cite{de2005tutorial} for code measures the naturalness of code against the code language model~\cite{hindle2016naturalness}.
Very recently, researchers have shown that code naturalness in terms of entropy values computed by LLMs can help rank patches for faster program repair~\cite{xia2022less, kolak2022patch}. Furthermore, Xia~\etal ~\cite{xia2022practical} demonstrated for the first time that it is possible to use such entropy values to perform patch correctness checking, i.e., \textit{correct patches can be more natural than overfitting patches.}\Comment{Kolak~\etal~\cite{kolak2022patch} found that larger models tend to consider developer patches more natural. Xia~\etal~\cite{xia2022practical} proposed to use entropy to represent code naturalness of generated patches and rank patches.} However, there is no comprehensive study on how such entropy-based techniques perform in \pcc{} datasets to our knowledge.\Comment{ LLM entropy-based techniques are based on the assumption that \textit{correct patches should be more natural than overfitting patches.}} 
For a list of tokens in the corpus of LLM\Comment{\lingming{pls don't say generation (which may confuse the reviewers), there is NO GENERATION, and we just use LLM to compute the naturalness}}, the mean entropy can be calculated as the negative log probability of each token as $mean\_entropy=-\sum_{i=1}^n\frac{log(p_{t_i})}{n}$, where $t_i$ refers to the ith token of the sequence and $p_{t_i}$ refers to the model probability of token $t_i$. 
Similarly, the sum entropy can be computed as $sum\_entropy=-\sum_{i=1}^nlog(p_{t_i})$. 
\Comment{shown below ($t_i$ refers to the ith token of the sequence and $p_{t_i}$ refers to the model probability of token $t_i$):
\begin{center}
    \begin{equation}
        mean\_entropy=-\sum_{i=1}^n\frac{log(p_{t_i})}{n}
    \end{equation}
    \begin{equation}
        sum\_entropy=-\sum_{i=1}^nlog(p_{t_i})
    \end{equation}
\end{center}}

Patches with lower sum or mean entropies are considered more \textit{natural} and will be ranked higher. In our evaluation, we use the model CodeT5-large~\cite{wang2021codet5}, and follow the same experimental setting with prior work~\cite{xia2022practical}.\Comment{ For a given patch, we 1)extract and tokenize the context of the patch to context tokens, 2)tokenize the target patch to target tokens, 3) iteratively generate the target tokens, get the probability of tokens one by one and append the tokens to the context tokens (before patch) for next iteration (one at a time).} For entropy-based techniques, we only use \avr(as described in section~\ref{sec:setup:tech:static}) to evaluate the effectiveness of patch ranking since the entropy is dependent on the context of bugs and improper for comparing patches across different bugs.


\subsubsection{Dynamic Techniques}
Dynamic techniques are designed to capture testing behavior and result for patch assessment.
\Comment{Both the correct and overfitting patches can pass the original provided test suite since they are all plausible ones. 
Therefore, new tests are required to be generated to differentiate correct ones from overfitting ones.}
Specifically, we focus on state-of-the-art dynamic techniques widely studied in the literature~\cite{yang2017better,DBLP:conf/kbse/WangWLWQZMJ20,xiong2018identifying}:

\textbf{Opad:} Opad is designed based on the hypothesis\Comment{\lingming{change into `hypothesis' to avoid confusion with oracle-based pcc} \jun{fixed}} that \textit{patches should not introduce new crash or memory-safety problems}~\cite{yang2017better}. 
Therefore, any patch violating those rules is regarded as overfitting.
Note that Opad is originally designed for C programs utilizing fuzzing techniques to generate new test cases. 
To adapt it on Java, we leverage two state-of-the-art test generation tools, i.e., \textit{Evosuite}~\cite{Fraser2011EvoSuite} and \textit{Randoop}~\cite{pacheco2007randoop}, to generate test cases based on the buggy version.
Opad based on \textit{Evosuite} and \textit{Randoop} are denoted as E-Opad and R-Opad respectively, following the previous study~\cite{DBLP:conf/kbse/WangWLWQZMJ20}.
{Specifically, for \textit{Evosuite} and \textit{Randoop}, we generate 30 test suites for each buggy program with a time budget of 600 seconds for each test suite utilizing existing test generation module in \dfj framework. We finally successfully generate 44,937 test suites on 796 buggy programs in total.}
After test generation, we first run the generated tests (for five times) on the original buggy programs to remove flaky tests.
After that, we run the remaining ones on the patched programs.
If any crash occurs or any exception is thrown, Opad will identify the patch as overfitting, otherwise it will identify the patch as correct.

\textbf{\patchsim:} \patchsim is a similarity-based \pcc{} tool, which utilizes the tests generated by automated test generation tools (Randoop in the original study~\cite{xiong2018identifying} and both Randoop and Evosuite in Wang \etal's study~\cite{DBLP:conf/kbse/WangWLWQZMJ20}) as the test inputs. It assumes that tests with similar executions are likely to have similar results. 
Therefore, given that a correct patch may behave similarly on the passing tests and differently on the failing ones compared with the buggy program, \patchsim could assess patch correctness\Comment{ of a patch based on these heuristics. a little confused here, which heuristics?}. 
Note that in the original study, Xiong \etal~used Randoop to generate test suites~\cite{xiong2018identifying}.
Besides, according to Wang \etal's study~\cite{DBLP:conf/kbse/WangWLWQZMJ20}, the tests generated by Randoop perform better than those by Evosuite.
We, thus, choose Randoop as the test generation tool in this study. To apply \patchsim on \dfjoldprapr, we generate 6,570 test cases for related projects in total by ourselves. Combining with the test suites generated by the existing study~\cite{DBLP:conf/kbse/WangWLWQZMJ20}, we obtain 10,404 test cases for \patchsim. 

\subsubsection{Learning-based Techniques}
Besides the traditional \pcc{} tools, we also include novel learning-based technique with different embedding techniques for comparison (hereafter denoted as  \textit{Embedding}), in our study. 
In addition, we include ODS proposed by Ye~\etal~\cite{ye2021ods} which aims to learn patch correctness via feature engineering. Lin~\etal~\cite{lin2022context} proposed Cache with similar ideas but their tools are currently not executable due to broken scripts and incomplete artifacts. 

\textbf{Embedding:} 
\textit{Code Embedding} is the technique that can transfer source code into distributed representations as fixed-length vectors. Consequently, various software engineering tasks can take the advantages of such distributed representations by utilizing supervised machine learning algorithms.
Tian~\etal~initially proposed utilizing code embedding techniques to identify correct patches among plausible ones~\cite{tian2020evaluating}.
There are also other techniques utilizing embedding to classify patches~\cite{embedding_js_2021, codeembedding}, but they mostly follow similar ideas and cannot outperform Tian~\etal~'s work~\cite{tian2020evaluating}.\Comment{but the dataset they used is relatively small (with only 64 patches from Quixbugs~\cite{quixbugs} and 465 patches from BugsJS\cite{bugsjs}, individually) and they could not outperform Tian \etal.} \Comment{\lingming{just claiming hte datasets are outdated is insufficient. can they outperform tian?}. \yuehan{Added simple summary here.}}\Comment{\yuehan{Add another SOTA technique using embedding here.}\lingming{why Xiong's study is relevant here?}\yuehan{Typo, fixed.}} 
Therefore, we utilize the best-performing model in Tian~\etal~'s study, denoted as BERT-LR (i.e., the pre-trained models of BERT~\cite{DBLP:journals/corr/abs-1810-04805} with logistic regression), to investigate the performance of embedding-based technique. We follow the same setting as adopted in the existing study by utilizing the dataset used in~\cite{tian2020evaluating}~for training and our collected benchmark for testing. Note that we filter the overlapped patches  for the training process.


\textbf{ODS:}
\textit{ODS} (\textbf{O}verfitting \textbf{D}etection \textbf{S}ystem) extracts a large number of static code features and leverages ensemble learning based on decision trees to predict patch correctness~\cite{ye2021ods}. For a given patch, ODS extracts 202 code features, including 150 code description features representing the characteristics of the patch's ingredients and its context at different granularity, 26 repair pattern features encoded with human knowledge on repair strategies and 26 contextual syntactic features that encode the scope and similarity information in the source code. 
To evaluate ODS in our dataset, we follow the experimental setup in the original paper~\cite{ye2021ods}. We first use the code analyzer provided by ODS to extract features of patches in our dataset and then reuse the model trained by the authors for prediction.

\subsection{Metrics}~\label{sec:metric}
Following the recent works on patch correctness checking~\cite{DBLP:conf/kbse/WangWLWQZMJ20,xiong2018identifying,yang2017better,tian2020evaluating}, we evaluate \pcc{} techniques against following metrics:
\begin{itemize}[itemindent=0pt, topsep=0pt, leftmargin=*]
    \item TP: \# of truly \Comment{\lingming{`precisely labeled' seems strange, change into `truly' or better ones for all similar sentences} \jun{fixed}}overfitting patches identified as overfitting. 
    \item TN: \# of truly correct patches identified as correct. 
    \item FP: \# of truly correct patches identified as overfitting.
    \item FN: \# of truly overfitting patches identified as correct.
    \item Precision: $= TP/(TP+FP)$. Denoted as \overfitprecision.
    \item Recall: $= TP/(TP+FN)$. Denoted as \overfitrecall.
    \item Recall of Correct: $= TN/(TN+FP)$. Denoted as \correctrecall.\Comment{\lingming{why font for recallc is different? We shall also capitalize the first letter for Precision and Recall} \jun{fixed}}
    \item \prauc : \textbf{A}rea \textbf{U}nder \textbf{P}recision-\textbf{R}ecall \textbf{C}urve~\cite{davis2006relationship}.
    
    \item \avr: the \textbf{AV}erage \textbf{R}anking of correct patches for bugs which have both overfitting and correct patches.
    \item Correct patch ratio: $= TN/(TN+FN)$. Denoted as \cpr.
    \item Accuracy: $=(TP+TN)/(TP+FP+TN+FN)$. Denoted as \acc.
    
\end{itemize}

Studies suggest that \pcc{} techniques should not mistakenly exclude correct patches~\cite{DBLP:conf/kbse/WangWLWQZMJ20,xiong2018identifying, ye2021ods}, i.e., the FP should be as low as possible.\Comment{Patch-sim paper stated that we should identify as many overfitting patches as possible, even at the risk of missing some correct patches.} However, when the dataset is imbalanced (\ie, mostly overfitting patches), \overfitprecision can be biased and overrated since $FP$ is often much smaller than $TP$, thus posing significant effects to the final results. Therefore, following the previous work on imbalanced datasets~\cite{xiong2018identifying}, we further include the metric \correctrecall (the ratio of correctly identified correct patches\Comment{among all the correct ones}) for complementary fair comparisons. In fact, \correctrecall{} is crucial since rejecting a correct patch may cause a bug to be unfixed. Besides, We also utilize the \prauc ~\cite{davis2006relationship} metric to eliminate the effects caused by imbalanced datasets~\cite{saito2015precision}.
\Comment{\yuehan{Specifically, \prauc can provide better evaluation when the dataset is heavily imbalanced.}}

Note that \prauc is not applicable for rule-based techniques including Opad, \patchsim, \antipattern{} and ODS\Comment{because they classify patches based on certain rules instead of calculated values.} since the results are discrete (either 0 or 1). 
For the static techniques (S3, CapGen, ssFix and code naturalness) based on feature score ranking, we leverage \avr{} to show their average ranking of correct patches, i.e.,  the smaller ranking is better. For example, given \avr{} \textit{a(b)}, \textit{a} is the average ranking of the first correct patch in each bug while \textit{b} is the average number of  patches in each bug.
In summary, an effective \pcc{} technique should achieve high \overfitprecision{}, \overfitrecall{}, \correctrecall{} and \prauc{}, while small \avr{}. For ODS evaluation we reuse their metrics with two additional ones \cpr and \acc.

\Comment{\subsection{Experimental Procedure}~\label{sec:setup:procedure}
Especially showing the training/testing set construction for learning-based techniques.}

\subsection{Threats to Validity}

The threats to \emph{internal} validity mainly lie in the implementation of studied \pcc techniques. To reduce such threats, we reuse existing implementations whenever possible, and have carefully reviewed all our code and script. In addition, to mitigate the bias in the patch correctness labeling, we involve multiple participants, follow widely-used criteria summarized from existing work~\cite{DBLP:conf/icse/0001WKKB0WKMT20}, and have released all our patches for public review.
To mitigate the potential bias of randomness in sampling, we repeat the sampling process for ten times and use the average results.
The threats to \emph{external} validity are mainly concerned with the generalizability of our findings. 
We mitigate the threats from single APR tool by further merging our datasets with previous datasets.
\Comment{\yiling{this setence could also be shortened by saying:  we mitigate the threats from single APR tool by further merging our datasets with previous datasets.}}
The threats to \emph{construct} validity mainly lie in the metrics used in our study. To mitigate the threats, we have included all popular metrics for \pcc, including \overfitprecision, \overfitrecall, \correctrecall, \avr, \prauc, \cpr and \acc.

\section{Result Analysis}



\subsection{RQ1: Static Techniques}~\label{sec:rq2}

\subsubsection{Static Code features} \Comment{To fully understand how those existing static code features perform on our new datasets, we follow a recent study\cite{DBLP:conf/kbse/WangWLWQZMJ20} to evaluate and compare the results achieved by different sets of static code features proposed by existing techniques.
Our empirical results are as follows:}For studying all the static code features, we follow the same experimental setting as the recent study~\cite{DBLP:conf/kbse/WangWLWQZMJ20}. The detailed experimental results for the three studied techniques are shown in \Comment{Tables~\ref{tab:ssFix_performance}, \ref{tab:s3_performance} and \ref{tab:capgen_performance}, respectively. } Table~\ref{tab:static_performance}.
From the table, we can observe that compared with original scores on \shangwendata{}, the \overfitrecall{} scores all increase while the \correctrecall scores all substantially decrease \Comment{\lingming{decrease or increase compared to what? not clear} \jun{fixed}} on the \dfjoldprapr{} and \dfjnewprapr{} datasets. One direct reason is that the two new datasets are extremely imbalanced (with mostly overfitting patches). Meanwhile, the extremely low \correctrecall{} scores (i.e., only 5\%-20\% of correct patches are identified as correct)
\Comment{\lingming{add some basic explanation for \correctrecall{} within the parenthesis (e.g., the ratio of correct patches that ...); otherwise it is hard to follow} \jun{fixed}} still demonstrate that such techniques can hardly be useful in practice. 
The \prauc scores of all features for all datasets are below or around 50\%. 
Taking \dfjbalance dataset for example, two of the features get a \prauc score slightly over 50\% (the proportion of positive sample and the expected \prauc for random classification model) and one of them gets less than 50\%, indicating that these features are hardly useful to identify overfitting patches.
The experimental results on the more balanced \dfjmerge{} and \dfjbalance{} datasets further confirm this observation. 
For example, on the balanced dataset \dfjbalance{}, the \correctrecall{} remains similar to the prior study on \shangwendata~\cite{DBLP:conf/kbse/WangWLWQZMJ20} (i.e., $\sim$45\%), while the \overfitrecall{} substantially drops over 30 percentage points for all three static techniques (e.g., 78.7-46.5\% for \ssfix). 
\avr is a better metric for simulating actual efforts of selecting one correct patch from plausible patches. From the tables, we can observe that for \shangwendata dataset, developers would examine 2.1$\sim$2.8 patches on average (depending on the chosen technique) until the first correct patch is found. However, when \prapr dataset is considered or merged with \shangwendata, the AVRs rise to 5.5$\sim$8.2. In other words, when larger patch space is included, effectiveness of all similarity-based static tools significantly degrades and the cost to identify correct patch for each bug significantly increases.
%

\Comment{From the tables, we observe that the precision and recall vary from less than 50\% to over 95\% across different datasets.
The precision and recall on our datasets are close to ROP,\footnote{Ratio of postive instances in a dataset, here this refers to the percentage of overfitting patches in a dataset.}, while in \shangwendata they are about 6\% higher than ROP, that is to say these techniques are less effective on our datasets.
\civi{Why we cannot conclude if they are close ROP?}
When the \prapr patches are included, none of the three tools achieve \correctrecall higher than 50\%, and the AUC score declined by 5.99\% $\sim$ 11.47\% for ssFix, 6.48\% $\sim$ 19.46\% for S3, and 14.7\% $\sim$ 17.51\% for CapGen.
As for \correctrecall, we observe that the score of CapGen is the highest while that of ssFix is the lowest among all the four datasets.
Unfortunately, none of them can achieve \correctrecall higher than 50\% in any of our datasets.
To conclude, either in the pure \prapr dataset, \ie, \dfjoldprapr~and \dfjnewprapr~or a balanced dataset combining \prapr patches and the patches generated from 21 tools, \ie \dfjmerge and \dfjbalance, the effectiveness of the existing three static tools is poor.
Such results indicate that existing similarity-based static techniques are actually ineffective in distinguishing overfitting patches from correct ones. 
Therefore, we make the following conclusion:}

\Comment{

We first compare the results of all three static techniques on the aforementioned four datasets.
Table~\ref{tab:ssFix_performance}, Table~\ref{tab:s3_performance} and Table~\ref{tab:capgen_performance} show the results of three static techniques seperately.}

%

%

%

\begin{table}[h]
  \setlength\tabcolsep{2.0pt} 
  \def\arraystretch{0.8}
  \scriptsize
  \caption{Performance of static features on different datasets} 
  \centering
  \begin{tabular}{l|r|r|r|r|r|r|r|r|r|r}
\toprule
{} & Dataset & \textit{TP} & \textit{FP} & \textit{TN} & \textit{FN} & \textit{\overfitprecision} & \textit{\overfitrecall} & \textit{\correctrecall} & \textit{\prauc} & \textit{\avr}\\
\midrule
\multirow{5}{*}{S3} & \shangwendata & 517 & 137 & 111 & 137 & 79.1\% & 79.1\% & 44.8\% & 45.3\% & 2.8(9.1)\\
    & \dfjoldprapr & 1222 & 42 & 5 & 42 & 96.7\% & 96.7\% & 10.6\% & 4.1\% & 8.2(10.7)\\
    & \dfjnewprapr & 1826 & 79 & 4 & 79 & 95.9\% & 95.9\% & 4.8\% & 5.0\% & 7.2(9.6)\\
    & \dfjmerge & 2250 & 239 & 32 & 239 & 90.4\% & 90.4\% & 11.8\% & 15.3\% & 7.0(12.9)\\
    & \dfjbalance & 120 & 151 & 120 & 151 & 44.3\% & 44.3\% & 44.3\% & 51.1\% & -\\
\midrule

\multirow{5}{*}{ssFix} & \shangwendata & 515 & 139 & 109 & 139 & 78.7\% & 78.7\% & 44.0\% & 43.8\% & 2.7(9.1)\\
    & \dfjoldprapr & 1221 & 43 & 4 & 43 & 96.6\% & 96.6\% & 8.5\% & 8.7\% & 6.8(10.7)\\
    & \dfjnewprapr & 1826 & 79 & 4 & 79 & 95.9\% & 95.9\% & 4.8\% & 6.7\% & 6.3(9.6)\\
    & \dfjmerge & 2242 & 247 & 24 & 247 & 90.1\% & 90.1\% & 8.9\% & 10.1\% & 6.1(12.9)\\
    & \dfjbalance & 126 & 145 & 126 & 145 & 46.5\% & 46.5\% & 46.5\% & 48.3\% & -\\
\midrule

\multirow{5}{*}{CapGen} & \shangwendata & 510 & 144 & 104 & 144 & 78.0\% & 78.0\% & 41.9\% & 46.2\% & 2.1(9.1)\\
& \dfjoldprapr & 1222 & 42 & 5 & 42 & 96.7\% & 96.7\% & 10.6\% & 14.9\% & 7.4(10.7)\\
& \dfjnewprapr & 1827 & 78 & 5 & 78 & 95.9\% & 95.9\% & 6.0\% & 11.4\% & 6.6(9.6)\\
& \dfjmerge & 2253 & 236 & 35 & 236 & 90.5\% & 90.5\% & 12.9\% & 16.9\% & 5.5(12.9)\\
& \dfjbalance & 133 & 138 & 133 & 138 & 49.1\% & 49.1\% & 49.1\% & 54.8\% & -\\
\midrule
\end{tabular}
\label{tab:static_performance}
  \label{tab:static_performance}
  
\end{table}

With the aim to understand the reasons behind these results, we further investigate the detailed raw patch scores computed by different tools as shown in Table~\ref{tab:average_score}.
Note that for \ssfix and \capgen, the scores reveal the \textbf{similarity} between the buggy code and the patches; while for \sthree, the scores represent the \textbf{edit distance} of the patches.
Such techniques are designed based on the widely-accepted assumption that \textit{correct patches should be more similar to the buggy code.} To investigate the validity of such a widely-accepted assumption, 
we split all the patches into four groups: \textit{\dfjoldprapr correct patches}, \textit{\dfjoldprapr~overfitting patches}, \textit{\shangwendata~correct patches} and  \textit{\shangwendata~overfitting patches}.
Furthermore, we also include the \textit{developer patches} for comparison.
As shown in Table~\ref{tab:average_score}, if we focus on the dataset of \shangwendata, the average scores of correct patches for ssFix and CapGen are 10.37\% and 76\% higher than the overfitting patches respectively. 
\begin{table}
  \setlength\tabcolsep{5.0pt} 
  \def\arraystretch{0.8}
  \scriptsize
  \caption{Average scores based on static features \Comment{\lingming{add two hlines to separate the rows into three blocks, i.e., prapr, wang, and developer}\yuehan{Done}}} 
  \centering
  \begin{tabular}{l|r|r|r}
\toprule
Patches & \textit{ssFix} & \textit{S3} & \textit{CapGen}\\
\midrule
\dfjnewprapr Correct & 1.52 & 10.60 & 0.37\\
\dfjnewprapr Overfitting & 1.56 & 15.73 & 0.41\\
\midrule
\shangwendata Correct & 1.49 & 11.71 & 0.44\\
\shangwendata Overfitting & 1.35 & 26.35 & 0.25\\
\midrule
Developer & 1.16 & 40.64 & 0.26\\
\bottomrule
\end{tabular}
\label{tab:scores_of_patches}

  \label{tab:average_score}
  
\end{table}
Besides, the average score of the correct patches for S3 is 55.56\% lower than of the overfitting patches.
A Mann-Whitney U Test~\cite{mann1947test} is conducted to evaluate the significance of the difference and it shows that the p-values are relatively $4.19*10^{-7}$, $3.00*10^{-11}$ and $9.72*10^{-9}$ for \ssfix, \capgen and \sthree respectively.
That is to say, correct patches in \shangwendata~in general share higher similarities with the buggy code and introduce fewer modifications than overfitting patches.
Such results actually support the widely-accepted assumption, and can also explain why these tools exhibit promising results on \shangwendata{}~\cite{DBLP:conf/kbse/WangWLWQZMJ20} and their original publications~\cite{le2017s3,DBLP:conf/icse/WenCWHC18,DBLP:conf/kbse/XinR17/ssfix}.

However, such an assumption might no longer hold on our datasets. 
Specifically, on \dfjnewprapr{}, for ssFix and CapGen, the average similarity scores of the overfitting patches are even 2.63\% and 10.81\% higher than those of the correct patches (the p-values are 0.070 and 0.106 respectively).
Moreover, if we compare with correct developer patches, the average score of \ssfix on overfitting patches in \shangwendata{} is also 14.66\% higher than that of developer patches (with p-value of $1.01*10^{-11}$) while average scores of \ssfix and \capgen on \dfjnewprapr overfitting patches are 34.48\% and 57.69\% higher than that of developer patches (with p-values of $1.42*10^{-54}$ and $5.53*10 ^{-12}$).
Such results are contradictory to those observed merely based on \shangwendata~\cite{DBLP:conf/kbse/WangWLWQZMJ20}. 
For S3, despite the results observed on \dfjnewprapr~share a similar trend with those on \shangwendata, we can still observe contradictory results if compared with developer patches.
Specifically, the average distance scores of S3 on developer patches are 54.23\% higher in \shangwendata~and 158.36\% higher in \dfjnewprapr{} than overfitting patches (p-values of $1.87*10^{-25}$ and $4.92*10^{-81}$). 
We also investigate the difference of 8 corresponding sub-features separately which show similar trend, presented on our website~\cite{replication}. \Comment{ Such results indicate that 1) the widely-accepted assumption no longer holds on our new dataset with more exhaustive patch generation, 2) the assumption will be even for future \apr techniques that can potentially produce more developer patches.} 

\Comment{Such results reveal that if evaluated on our large-scale dataset, the correct patches are not necessarily have higher similarities with the original buggy code, and might also introduce more modifications.\jun{do not necessarily?}
More interesting, the results indicate that developer patches actually share less similarities with the buggy code and introduce more modifications measured with respect to the existing static code features, and thus they are more likely to be classified as overfitting by these tools instead of correct ones.
However, the previous studies~\cite{DBLP:conf/kbse/WangWLWQZMJ20} reported promising results since, actually, the patches generated by \prapr and the other 21 \apr tools can only fix a small subset of the \dfj bugs with rather simple fixing attempts.
However,\lingming{two `However'. Also, we should make clear about two things: 1) even for auto-generated patches the assumption does not hold when considering more realistic datasets, 2) developer patches can make the results even worse (but we should also make clear that this guideline is just for future APR tools as current APR tools cannot generate such developer patches)} the majority of remaining bugs require much more complex patches to fix. Therefore, those patches generated by the existing APR tools are generally more simple and similar to buggy code snippets compared with developer patches considering those patches that cannot be generated by the existing APR techniques. To summary, we make the following conclusion:
}
\finding{The widely-accepted assumption made by existing similarity-based static techniques that correct patches in general share higher similarities with the buggy code is no longer valid on our new dataset with exhaustive patch generation.}
Through further investigation, we observe that one major issue of similarity-based static techniques is that they can produce diverse similarity scores for semantic-equivalent patches\Comment{ as long as they are syntactically different}.
\Comment{Consequently, they cannot precisely distinguish overfitting patches from correct ones as revealed by our previous results.}
For instance, Listing~\ref{lst:remove_if_block} shows a simple example patch.
\begin{listing}[]
\begin{minted}{java}
-   if (this.rightBlock != null) {
+   if (false) {
        RectangleConstraint c4 = new RectangleConstraint...
\end{minted}
        
        \caption{A \prapr patch replacing condition with \textit{false}}
        \label{lst:remove_if_block}
\end{listing}
\begin{table}[]
  \setlength\tabcolsep{4.0pt} 
  \def\arraystretch{0.8}
  \scriptsize
  \caption{Static PCC on semantically equivalent patches\Comment{\lingming{looks a bit ugly. transpose this table} \jun{fixed}}} 
  \centering

\begin{tabular}{l|r|r|r|r}
\toprule
Chart-13-mutant-6 & \textit{TokenStrct} & \textit{TokenConpt} & \textit{ASTDist} & \textit{ASTCosDist}\\
\midrule
mutate condition to false & 0.829 & 0.773 & 6 & 0.025\\
remove whole block & 0.344 & 0.419 & 51 & 0.097\\
\midrule
Chart-13-mutant-6 & \textit{VariableDist} & \textit{VariableSimi} & \textit{SyntaxSimi} & \textit{SemanticSimi}\\
\midrule
mutate condition to false & 3 & 0.5 & 0.245 & 0.011\\
remove whole block & 20 & 0.393 & 0.347 & 0.618\\
\bottomrule
\end{tabular}
\label{tab:scores_for_remove_block_or_not}
  
  \label{tab:scores_for_remove_block_or_not}
\end{table}
This patch simply replaces the conditional expression with \CodeIn{false}, which is semantically equivalent to removing the whole if block (including its large body).
The scores generated by the three static techniques are totally different, as shown in Table~\ref{tab:scores_for_remove_block_or_not}.
The left column displays the scores of the original patch, \ie, replace the conditional expression with \CodeIn{false}, while the right column denotes the scores of the simplified patch, \ie, simply removing the whole \textit{if} block.
Among the eight static features, the similarity scores of \ssfix features of original patch are significantly higher compared with the simplified patch (\ie, the patch with the whole \textit{if} block removed) 140.99\% and 84.49\% respectively, while the edit distance scores for \sthree of the original patch have much lower scores than the simplified patch. For \capgen features, the scores are also completely different from each other.
These cases are frequently observed in our dataset, \ie, a patch can be simplified to its equivalent patch. There are 54 \textit{mutating condition to false} patches among the 1,311 \dfjoldprapr patches, in addition to other patches with similar patterns.
Such results provide stronger evidence that merely considering the syntactic similarity is inadequate for \pcc~while more advanced semantics-based techniques should be proposed.

\finding{All three studied static techniques can compute totally different scores for semantically equivalent patches, indicating that future static \pcc techniques should incorporate more advanced semantics analysis.}

\begin{table}[h]
  \setlength\tabcolsep{4.0pt} 
  \def\arraystretch{0.8}
  \scriptsize
  \caption{Performance of \antipattern{}}
  \centering
  \begin{tabular}{l|r|r|r|r|r|r|r}
\toprule
Datasets & \textit{TP} & \textit{FP} & \textit{TN} & \textit{FN} & \textit{\overfitprecision{}} & \textit{\overfitrecall{}} & \textit{\correctrecall} \\
\midrule
\shangwendata & 219 & 37 & 211 & 435 & 85.55\% & 33.49\% & 85.08\% \\
\dfjoldprapr & 174 & 10 & 37 & 1,090 & 94.57\% & 13.77\% & 78.72\%\\
\dfjnewprapr & 361 & 28 & 55 & 1,544 & 92.80\% & 18.95\% & 66.27\%\\
\dfjmerge & 559 & 56 & 215 & 1,930 & 90.89\% & 22.46\% & 79.34\%\\
\dfjbalance & 41 & 56 & 215 & 230 & 42.27\% & 15.13\% & 79.34\%\\
\bottomrule
\end{tabular}
  
  \label{tab:anti_final}
\end{table}

\subsubsection{Anti-patterns}~\label{sec:res_anti}
Table~\ref{tab:anti_final} presents the results of \antipattern{}~\cite{Tan2016anti} on our four new datasets, along with its original results in previous work~\cite{DBLP:conf/kbse/WangWLWQZMJ20} (i.e., \shangwendata{}).  From the table, we can observe that \antipattern{} performs significantly worse on \prapr patches by misidentifying a larger proportion of \Comment{\yuehan{Should the word "more" be replaced with "larger proportion"? Considering that FP in PraPR is less than that in Wang's dataset} \jun{I think it's OK.}}correct patches as overfitting and omitting more overfitting patches. Compared to the previous dataset,  \antipattern{} exhibits both lower \correctrecall{} and \overfitrecall{} on \dfjoldprapr{}. Although \overfitprecision{} has a suspected improvement, the actual cause is the heavily imbalanced datasets, i.e., the overwhelming number of overfitting patches over the correct ones. Actually, the results on the balanced dataset \dfjbalance{} further confirm the worse performance. 

\Comment{\antipattern{} actually performs as a random approach on the balanced dataset (i.e., almost 50\% \overfitprecision{} and \overfitrecall{} on \dfjbalance{}). Such observations further indicate that  \antipattern{} is still far from being practical.}

We further looked into the root cause of \antipattern{}' poor performance on our datasets with two misclassified examples of \antipattern{}. Note that these two patches are generated exclusively by \prapr{} and have not been included in any previous dataset.
Listing~\ref{lst:anti_fp_3} is an FP example where \antipattern{} mistakenly identifies a correct patch generated by \prapr{} as overfitting. In particular, the correct patch shown in Listing~\ref{lst:anti_fp_3} falls in the anti-pattern A2 which disallows any removal of control statements such as if-statement. In fact, in order to enable general and efficient searching attempts in the patch space,  fixing operations leveraged in existing APR (especially the basic fixing operations in \prapr{}) are often much less sophisticated than manual fixes, which may even appear ``simple'' and ``crude'' sometimes. Therefore, \antipattern{}, which distinguishes overfitting and correct patches by the pattern of manual patches, inherently becomes less effective in identifying the developer-unlike\Comment{not sure whether this expression is correct} but correct patches (which are largely missed by prior \apr tools but captured by \prapr).
\Comment{Similarly, the correct patch in List~\ref{lst:anti_fp_2} falls in to another anti-pattern A5 (i.e., Anti-delete Loop-Counter). 
For any assignment statement within a loop, A5 disallows its deletion if its LHS uses variables in the terminating condition.} 

\begin{listing}[h]
\begin{minted}{java}
     if (childType.isDict()) {...
     } else {
-       if (n.getJSType() != null && parent.isAssign())
+       if (false && parent.isAssign())
          return;
     } else if ...
\end{minted}
        
        \caption{A correct patch misclassified  by \antipattern{}}
        \label{lst:anti_fp_3}
\end{listing}

\begin{listing}[h]
\begin{minted}{java}
-   double b = calculateBottomInset(h);
+   double b = h;
    area.setRect(area.getX() + l, 
    area.getY() + t, w - l - r, h - t - b);
\end{minted}
        
        \caption{An overfitting patch misclassified by \antipattern{}}
        \label{lst:anti_fp_4}
\end{listing}

Listing~\ref{lst:anti_fp_4} presents an FN example that \antipattern{} mistakenly identifies overfitting patches generated by \prapr{} as correct. The overfitting patch makes a subtle modification by removing the method invocation \texttt{calculateBottomInset} (i.e., only one token changed). Such a modification does not fall into any existing \antipattern{}, thus cannot be excluded during \pcc{}. In particular, although this overfitting patch involves only a basic repair operation, it is not included in any other existing dataset. This further confirms our motivation that the dataset-overfitting issue and early-stop mechanism of most APR tools result in insufficient plausible patches for evaluating \pcc{}\Comment{, and including patches generated by \prapr{} can help explore a larger plausible patch space due to its plentiful basic repair operations}.\Comment{ In addition, the clearly reduced effectiveness of \antipattern{} on \prapr{} datasets also indicates the necessity to evaluate  \pcc{} techniques in a larger plausible patch space.}
\finding{\antipattern{} performs much worse on our datasets due to its limited capability of identifying (i) overfitting patches with subtle modifications and (ii) correct patches with developer-unlike modifications. Such problem is largely amplified on our datasets due to \prapr's exhaustive patch generation.}

\subsubsection{\textit{Naturalness-based} techniques}
\begin{table}[h]
  \setlength\tabcolsep{4.0pt} 
  \def\arraystretch{1.0}
  \scriptsize
  \caption{Performance of code naturalness}
  \centering
  \begin{tabular}{l|r|r|r|r|r}
\toprule
\avr & \textit{ssFix} & \textit{s3} & \textit{capgen} & \textit{\sumentropy} & \textit{\meanentropy}\\
\midrule
\shangwendata & 2.72(9.1) & 2.8(9.1) & 2.14(9.1) & 1.9(9.1) & 1.9(9.1)\\
\dfjoldprapr & 6.77(10.69) & 8.15(10.69) & 7.38(10.69) & 4.73(10.69) & 4.65(10.69)\\
\dfjnewprapr & 6.26(9.64) & 7.15(9.64) & 6.62(9.64) & 4.08(9.64) & 4.67(9.64)\\
\dfjmerge & 6.12(12.87) & 7(12.87) & 5.49(12.87) & 3.07(12.87) & 3.47(12.87)\\
\bottomrule
\end{tabular}
\label{tab:entropy_rank}
  \label{tab:entropy_rank}
\end{table}
%

Table~\ref{tab:entropy_rank} shows that in all datasets, both \sumentropy and \meanentropy can outperform all three static techniques. Specifically, in terms of AVRs, the sum/mean entropy outperforms the best-performed static technique by 11.21\%/11.21\% (on \shangwendata), 30.13\%/31.31\% (on \dfjoldprapr), 34.82\%/25.40\% (on \dfjnewprapr), and 44.08\%/36.79\% (on \dfjmerge). Interestingly, our results also confirm prior work~\cite{xia2022practical} that \sumentropy performs slightly better than \meanentropy for PCC. The reason is that \sumentropy calculates the entire sequence entropy for code naturalness computation, and considers both code naturalness and length information.\Comment{ thus shorter sequences tend to get lower naturalness, which is consistent with traditional static techniques~\cite{cite s3, capgen, etc} (which also favor simple patches over complicated ones).}\Comment{relatively 11.21\% and 11.21\%(on \shangwendata), 30.13\% and 31.31\%(on \dfjoldprapr), 34.82\% and 25.40\%(on \dfjnewprapr), 44.08\% and 36.79\%(on \dfjmerge) higher than the best-performed static code features.} 
To conclude, \naturalness techniques show greater potential in \pcc and this is the first, to our best knowledge, evaluation of \naturalness techniques on \pcc datasets. We appeal to the community that more attention should be drawn into this direction.
\finding{\naturalness techniques can substantially outperform all static code features in patch ranking, and \sumentropy performs slightly better than \meanentropy.}

\subsection{RQ2: Learning-based Techniques}
\subsubsection{Embedding technique}
\begin{table}[]
  \setlength\tabcolsep{3.0pt} 
  \def\arraystretch{1.0}
  \scriptsize
  \caption{Performance of the embedding-based technique}
  \centering
  \begin{tabular}{l|r|r|r|r|r|r|r|r}
\toprule
Datasets & \textit{TP} & \textit{FP} & \textit{TN} & \textit{FN} & \textit{\overfitprecision{}} & \textit{\overfitrecall{}} & \textit{\correctrecall} & \textit{PR-AUC}\\
\midrule
\tiandata{} & 85 & 14 & 16 & 24 & 85.86\% & 77.98\% & 53.33\%  & N/A\\
\dfjoldprapr & 822 & 15 & 27 & 227 & 98.21\% & 78.36\% & 64.29\%  & 11.86\%\\
\dfjnewprapr & 1,220 & 23 & 53 & 433 & 97.52\% & 72.81\% & 59.21\% &  15.51\%\\
\dfjdelta & 398 & 8 & 26 & 206 & 96.14\% & 65.89\% & 52.94\% &  19.69\%\\
\bottomrule
\end{tabular}
  \label{tab:embedding_final}
\end{table}
Table~\ref{tab:embedding_final} presents the results of the embedding-based technique~\cite{tian2020evaluating} on our datasets, i.e., \dfjoldprapr{} and \dfjnewprapr{}. Since there is a large overlap between the training set of the embedding technique and \dfjmerge{}/\dfjbalance, i.e., patches collected by Liu~\etal{}~\cite{DBLP:conf/icse/0001WKKB0WKMT20}, we do not consider \dfjmerge{} or \dfjbalance{} in this RQ. For comparison, we also present the results of the  embedding-based technique in its original paper (i.e., \tiandata{}).

To evaluate the performance more precisely, we pay more attention to PR-AUC since the applicable datasets are heavily imbalanced. 
Obviously, the embedding-based technique tends to perform badly on \prapr{} patches with PR-AUC to be less than 20\% on  all our new datasets. Furthermore, with the datasets to be gradually more balanced through \dfjoldprapr{}, \dfjnewprapr{} and \dfjdelta{} (i.e., exclusive bugs in \dfjnewprapr{}), the performance of \overfitprecision{}, \overfitrecall{} and \correctrecall{} keeps dropping while the \prauc{} increases, which indicates the worse performance is more credible.  The results further confirms that the embedding-based technique suffers from the dataset overfitting issue: it performs worse on the patches of subjects that are different from training set. Thus, we encourage future learning-based \pcc{} work to consider across-dataset evaluation.

\Comment{In particular, compared to original results, \auc{} becomes lower on both \dfjoldprapr{} and \dfjnewprapr{}. Note that although there is an improvement in \overfitprecision{} and \correctrecall{}, it does not necessarily indicate that the embedding-based technique becomes more effective, since our datasets are more imbalanced than previous ones so \prauc{} are more reliable metrics for such imbalanced datasets.
\Comment{\correctrecall{} substantially drops by almost 30\%, indicating that the embedding-based technique mistakenly identifies a non-trivial portion of correct patches generated by \prapr{} as overfitting. Although there is a slight improvement in \overfitprecision{} and \overfitrecall{}, it does not necessarily indicate that the embedding-based technique becomes more effective in identifying overfitting patches, since our datasets are more imbalance than previous. In fact, the decrements in \auc{} show that  embedding-based technique has a weaker capability of identifying overfitting patches. 
}}

\Comment{\lingming{cross-subject or cross-dataset?} \yuehan{I prefer cross-dataset here, replaced for now.}} \Comment{ (i.e., the test set should include subjects absent from the training set)}  

\begin{figure}[]
    \centering
    \includegraphics[height= 3.5 cm]{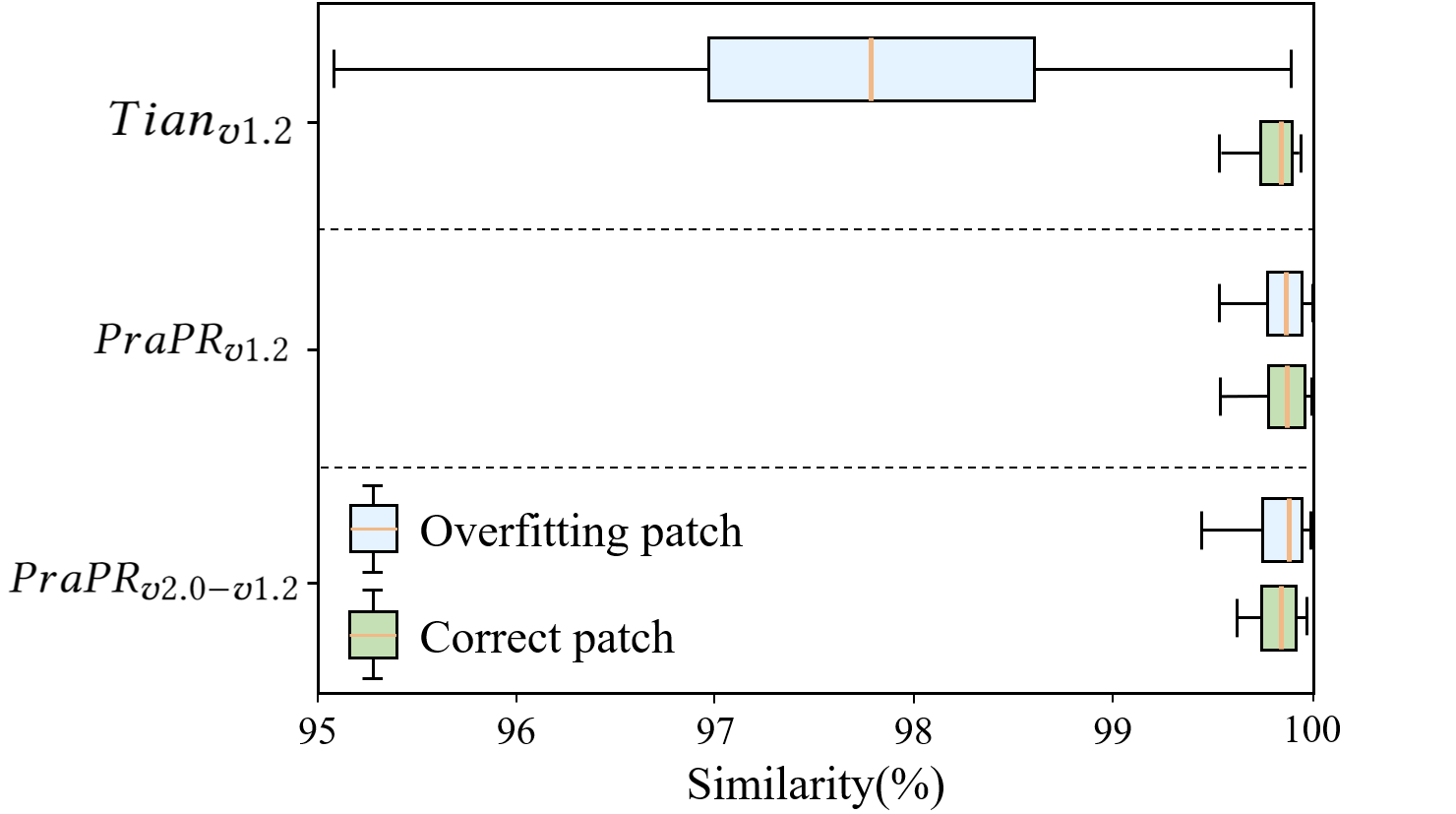}
    \caption{Similarity distribution on different datasets \Comment{\lingming{the style does not look quite good; please redraw}\yuehan{Done}}}
    
    \label{fig:embedding_cosine}
\end{figure}

We then look into the potential reasons for such a decrement. 
The intuition of embedding-based \pcc{} techniques is that the cosine similarity between the embedding vectors of correct patches and buggy code should be larger than the cosine similarity between the embedding vectors of  overfitting patches and buggy code. 
Tian~\etal{}~\cite{tian2020evaluating} show that correct/overfitting patches in their dataset perfectly follow such an assumption.
\Comment{
Tian~\etal{}~\cite{tian2020evaluating} show that the cosine similarity between the embedding vectors of correct/overfitting patches and buggy code perfectly follows such an assumption in their dataset.} However, this assumption no longer holds on the additional patches generated by \prapr{}. Figure~\ref{fig:embedding_cosine} presents the distribution of the cosine similarity between the embedding vectors of correct/overfitting patches and buggy code on the original dataset in the embedding work~\cite{tian2020evaluating} (i.e., \tiandata{}) and our datasets built on \dfjold{} and additional bugs in \dfjnew{} (i.e., \dfjoldprapr{} and \dfjdelta{}).
From the figure, we observe that different from the prior work~\cite{tian2020evaluating}, correct patches and overfitting patches share very close distributions of similarity scores on our datasets, which explains why the embedding-based technique exhibits worse performance on \prapr{} patches. Furthermore, correct patches even tend to have lower median similarity than overfitting patches on \dfjdelta{}, demonstrating that the assumption made by the embedding-based work no longer holds on our new dataset.
\finding{The assumption that the cosine similarity between the embeddings of correct patches and buggy code should be larger than that between the embeddings of overfitting patches and buggy code no longer holds.
\Comment{\yuehan{Can we shorten here with the content of the assumption to be removed?}} Also, the embedding-based technique tends to suffer from the dataset overfitting issue.  }

\Comment{Second, it is notable that the embedding-based technique performs even worse on \dfjnewprapr{} than \dfjoldprapr{}, i.e., \auc{} degrades by 3\% from \dfjoldprapr{} to \dfjnewprapr{}. It indicates that  embedding-based techniques are less effective on the additional bugs in \dfjnewprapr{}. To confirm the observation, we further present the results on the exclusive bugs in \dfjnewprapr{} in the table, i.e., \dfjdelta{}. The even lower \auc{} (i.e., 67.80
\%) indicates that embedding-based techniques suffer from an overfitting issue in datasets: it performs worse on the patches of subjects that are different from training set. Therefore, we encourage future learning-based \pcc{} techniques to consider across-subject evaluation (i.e., the test set should include subjects that are absent from the training set).  }

\subsubsection{ODS}
\begin{table}[h]
  \setlength\tabcolsep{3.0pt} 
  \def\arraystretch{1.0}
  \scriptsize
  \caption{Performance of ODS}
  \centering
  \begin{tabular}{l|r|r|r|r|r|r|r|r|r|r}
\toprule
ODS & \textit{TP} & \textit{FP} & \textit{TN} & \textit{FN} & \textit{precision} & \textit{recall} & \textit{\correctrecall} & \textit{F1} & \textit{CPR} & \textit{accuracy}\\
\midrule
\shangwendata & 620 & 66 & 182 & 34 & 90.38\% & 94.80\% & 73.39\% & 92.54\% & 84.26\% & 88.91\%\\
\dfjoldprapr & 1034 & 32 & 14 & 229 & 97.00\% & 81.87\% & 30.43\% & 88.79\% & 5.76\% & 80.06\%\\
\dfjnewprapr & 2571 & 93 & 35 & 592 & 96.51\% & 81.28\% & 27.34\% & 88.24\% & 5.58\% & 79.19\%\\
\bottomrule
\end{tabular}
\label{tab:ods}
  \label{tab:ods}
\end{table}
Table~\ref{tab:ods} shows the result of ODS in our dataset. For \shangwendata, we directly reuse the results from ~\cite{ye2021ods}.
It turns out that the performance on \dfjoldprapr{} and \dfjnewprapr{} significantly drop for \correctrecall{}, \cpr and \acc, \eg, \cpr drops from \textasciitilde84\% to\textasciitilde6\%, which means a lot of overfitting patches `escape' ODS and get misidentified as correct. 
Similarly, the \correctrecall{} drops because more correct patches are classified as overfitting. 
More interestingly, we evaluate ODS on developer patches and found that the FP rate (ratio of developer patches identified as overfitting) is 43.52\% and 41.56\% on developer patches v1.2 and v2.0. 
To figure out the potential reason that the performance drops, we carefully check the feature definitions in ODS and understand why \prapr{} patches can escape ODS. 
ODS defines 202 features in total (in three groups). According to the feature analysis of ODS~\cite{ye2021ods}, the order of importance of three groups of features is: Contextual Syntactic Features (150) > Code Description Features (26) > Repair Pattern Features (26). Note that ODS features are  specific characteristics of programs like whether the patch variables are local, global, primitive, \etc, or whether there's code move. 
For single-line replacement repair tool \prapr{}, it normally makes simple change like replacing a variable/constant/method with another one of the same type according to the mutators defined in ~\cite{DBLP:conf/issta/GhanbariBZ19}. Therefore, most \prapr{} patches do not affect the majority of the ODS features at all.
Listing~\ref{lst:chart_ods_escape} is an example patch for bug Chart-26 showing how \prapr{} patches escape ODS. \prapr{} generates 103 patches for this bug in total and 23 of them are classified as correct by ODS (while actually only two of them are correct). 11 of them simply replace \CodeIn{calculateLeftInset} with other APIs and are all identified as correct by ODS. Such cases produce large amount of FNs and thus lead to very low CPR(\textasciitilde6\%).
\begin{listing}[h]
\begin{minted}{java}
-    double b = calculateBottomInset(h);
+    double b = trimWidth(h);
\end{minted}
        
        \caption{An overfitting patch that escapes ODS}
        \label{lst:chart_ods_escape}
\end{listing}

\finding{Features of ODS are too weak to identify overfitting patches and miss a lot of overfitting patches in our dataset. Plus, the high FP rate on developer patches is unbearable for \pcc.}

\subsection{RQ3: Dynamic Techniques}~\label{sec:rq3}

\subsubsection{Opad}
\Comment{We performed the following quantitative analysis and qualitative analysis to investigate the effectiveness of Opad in separate.

\textbf{Quantitative Analysis.}} 
Table~\ref{tab:E_opad_performance} and Table~\ref{tab:R_opad_performance} respectively show the results for E-Opad (Opad with EvoSuite) and R-Opad (Opad with Randoop).
We can observe that except for R-Opad on \dfjbalance, Opad achieves a precision over 95\% on all the datasets, while the achieved recall ranges from 11.71\% to 18.83\%.
\begin{table}[h]
  \setlength\tabcolsep{4.0pt} 
  \def\arraystretch{1.0}
  \scriptsize
  \caption{Performance of E-Opad on different datasets} 
  \centering

\begin{tabular}{l|r|r|r|r|r|r|r}
\toprule
E-Opad & \textit{TP} & \textit{FP} & \textit{TN} & \textit{FN} & \textit{\overfitprecision{}} & \textit{\overfitrecall{}} & \textit{\correctrecall}\\
\midrule
\shangwendata & 92 & 0 & 248 & 562 & 100.00\% & 14.07\% & 100.00\%\\
\dfjoldprapr & 148 & 0 & 47 & 1,116 & 100.00\% & 11.71\% & 100.00\%\\
\dfjnewprapr & 267 & 2 & 81 & 1,638 & 99.26\% & 14.02\% & 97.59\%\\
\dfjmerge & 344 & 2 & 269 & 2,145 & 99.42\% & 13.82\% & 99.26\%\\
\dfjbalance & 48 & 2 & 269 & 223 & 96.00\% & 17.71\% & 99.26\%\\
\bottomrule
\end{tabular}
\label{tab:E_opad_performance}
  \label{tab:E_opad_performance}
  
\end{table}
\begin{table}[h]
  \setlength\tabcolsep{4.0pt} 
  \def\arraystretch{1.0}
  \scriptsize
  \caption{Performance of R-Opad on different datasets} 
  \centering

\begin{tabular}{l|r|r|r|r|r|r|r}
\toprule
R-Opad & \textit{TP} & \textit{FP} & \textit{TN} & \textit{FN} & \textit{\overfitprecision{}} & \textit{\overfitrecall{}} & \textit{\correctrecall}\\
\midrule
\shangwendata & 67 & 0 & 248 & 587 & 100.00\% & 10.24\% & 100.00\%\\
\dfjoldprapr & 238 & 12 & 35 & 1,026 & 95.20\% & 18.83\% & 74.47\%\\
\dfjnewprapr & 346 & 15 & 68 & 1,559 & 95.84\% & 18.16\% & 81.93\%\\
\dfjmerge & 408 & 15 & 256 & 2,081 & 96.45\% & 16.39\% & 94.46\%\\
\dfjbalance & 44 & 15 & 256 & 227 & 74.58\% & 16.24\% & 94.46\%\\
\bottomrule
\end{tabular}
\label{tab:R_opad_performance}
  \label{tab:R_opad_performance}
  
\end{table}
The high precision with low recall achieved by Opad is consistent with previous studies~\cite{DBLP:conf/kbse/WangWLWQZMJ20, yang2017better}.
Besides, the \correctrecall also significantly outperforms other techniques with respect to precision.
Such results indicate that Opad can identify most of the correct patches and rarely classifies correct patches as overfitting ones.
\Comment{Different from the static and learning-based techniques, the performance of Opad does not degrade much in our new datasets.} \Comment{\civi{not degraded under what criteria?}}
In other words, the performance does not degrade much in our new datasets, and Opad is more stable than static techniques.
This falls into our intuition since such dynamic tools based on test generation concern more towards code semantics instead of syntactic elements.
Meanwhile, similar to the findings from prior work~\cite{DBLP:conf/kbse/WangWLWQZMJ20}, Opad achieves a rather low recall on our datasets, indicating that a substantial ratio of overfitting patches are not detected.
This is still a critical issue in \pcc{} since manually filtering out overfitting patches can be extremely costly for developers~\cite{xiong2018identifying}.
\finding{The performance of Opad overall tends to remain similar on our new datasets, demonstrating the robustness of such dynamic techniques\Comment{does not degrade like other techniques on our new datasets}. Meanwhile, consistent with the findings in prior work, it achieves a rather low recall for incorrect patches\Comment{\lingming{for correct or incorrect patches?}\jun{Incorrect ones, the positive instance we defined}}, which compromises its practical usefulness.}
Though the precision achieved by Opad is still high in our new datasets, it can no longer achieve 100\% preision, i.e., a few correct patches are misidentified as overfitting.
This is inconsistent with the results \option{released }in previous studies for \pcc~\cite{DBLP:conf/kbse/WangWLWQZMJ20,yang2017better}\option{, where a precision of 100\% can be always achieved by Opad}.
Motivated by this, we further investigated the FP cases and made the following  observations.
%

%
\begin{listing}[h]
\begin{minted}{java}
-  if (property == null) {
-      return this;
-  }
   JsonFormat.Value format = findFormatO(serializers,property,handledType());...
\end{minted}
        
        \caption{A correct patch misclassified by \opad }
        \label{lst:Opad_prapr_false_positive_patch}
\end{listing}

For instance, Listing~\ref{lst:Opad_prapr_false_positive_patch} shows a correct patch which is identified as overfitting by Opad.
Specifically, this patch
removes a \CodeIn{null} check for \CodeIn{property}.
However, Opad is designed based on the tests generated on the original buggy programs, thus being unaware of such semantic changes.
When stepping into method \CodeIn{findFormatO} (at line 6) from a test, \CodeIn{property} is null, which is unexpected to the tests generated on the original buggy program.
Consequently, a \CodeIn{NullPointerException} was thrown and in \CodeIn{findFormatO}, and thus Opad mis-identified this patch as overfitting. 
%
%
Many other similar cases have been observed for Opad.
The reason could be that some tests are generated on the buggy programs, based on some incorrect contracts or preconditions.
When the semantics of a program change, those preconditions may not hold and thus fail a generated test.
\finding{The effectiveness of Opad might decay and it might mistakenly identify correct patches as incorrect when the semantic changes of the patches break certain conditions.}
\begin{table}[b!]
  \setlength\tabcolsep{10.0pt} 
  \def\arraystretch{0.6}
  \scriptsize
  \caption{\opad{}/\patchsim{} on developer patches\Comment{\lingming{add hline above patch-sim to separate it from the other two rows}\yuehan{Done}}} 
  \centering
  \begin{tabular}{l|r|r|r|r|r}
\toprule
 & \textit{TP} & \textit{FP} & \textit{TN} & \textit{FN} & \textit{\correctrecall}\\
\midrule
E-Opad & 0 & 111 & 685 & 0 & 86.06\%\\
R-Opad & 0 & 167 & 629 & 0 & 79.02\%\\
\midrule
\patchsim{} & 0 & 65 & 110 & 0 & 62.86\% \\
\bottomrule
\end{tabular}
  \label{tab:dev_on_dynamics}
  
\end{table}

Since developer patches might even incur larger semantic changes, we are curious to see whether such similar cases could also happen on developer patches.
Unfortunately, there are no existing studies investigating Opad on developer patches to our best knowledge. However, this is an important study since \apr techniques can potentially fix more and more bugs and produce more and more developer patches in the near future. As shown in Table~\ref{tab:dev_on_dynamics}, E-Opad and R-Opad achieve \correctrecall of 86.06\% and 79.02\%, respectively, which is significantly lower than that on the other datasets except for \dfjoldprapr.
%
Specifically, Opad produces more false-positives and achieves lower \correctrecall compared with previous study~\cite{DBLP:conf/kbse/WangWLWQZMJ20,yang2017better}.
We further manually checked some false positive cases in developer patches, and Listing~\ref{lst:Opad_developer_false_positive_patch} shows an example.
Specifically, the developer patch moved a statement into an \CodeIn{else} block with a condition \CodeIn{length >= 9}.
The failing test\Comment{, as shown in Listing~\ref{lst:Opad_developer_failing_test},}\Comment{ passed four parameters into method \CodeIn{TarUtils.formatLongOctalOrBinaryBytes}, \ie, a large long integer, a new byte array and two -1.
The failing test} expects an \CodeIn{IllegalArgumentException} triggered by invalid parameters of \CodeIn{formatBigIntegerBinary}.
When stepping into the method in Listing~\ref{lst:Opad_developer_false_positive_patch}, the original statement throwing that exception was not executed in the patched program.
Therefore, the program stepped over that statement and when executing the next statement (line 7)\Comment{\yuehan{Adjusted}}, the parameter \CodeIn{offset} is -1, triggering an \CodeIn{ArrayIndexOutOfBoundsException}.
\begin{listing}
\begin{minted}{java}
@@ -483,9 +483,8 @@ public static int formatLongOctalOrBinaryBytes(
        ...
         if (length < 9) {...
+         } else {
+            formatBigIntegerBinary(value, buf, offset, length, negative);}
-        formatBigIntegerBinary(value, buf, offset, length, negative);
         buf[offset] = (byte) (negative ? 0xff : 0x80);
         return offset + length;
\end{minted}
        
        \caption{A developer patch  misclassified by \opad}
        \label{lst:Opad_developer_false_positive_patch}
\end{listing}
\Comment{
\begin{listing}
\begin{minted}{java}
@Test(timeout = 4000)
public void test23()  throws Throwable  {
  byte[] byteArray0 = new byte[18];
  try { 
    TarUtils.formatLongOctalOrBinaryBytes(1099511627776L,byteArray0,(-1),(-1));
    fail("Expecting exception: IllegalArgumentException");
  } catch(IllegalArgumentException e) {
      verifyException("org.apache.commons.compress.archivers.tar.TarUtils",e);
  } }
\end{minted}
        
        \caption{Failing test of the developer patch}
        \label{lst:Opad_developer_failing_test}
\end{listing}}
%
Similar to the previous example, Opad is not aware of such semantic changes and misclassified that patch as overfitting.
Actually, Opad could produce more false-positives when the patches get more complicated, \ie, different code structure or logic. This can be an important finding for the community since eventually \apr techniques will be conquering more and more complicated bugs, with more and more complicated patches.
\Comment{This can be an important finding for the community because existing \apr techniques can only generate a limited number of relatively simple patches, and thus the evaluation based on the existing patches can be biased for Opad (which tends to perform worse on more complicated patches, which \apr tools can eventually cover). }
\Comment{\lingming{developer patches cannot be handled well by dynamic techs. this should provide future guidelines: the current techniques can only deal with simple changes well; while later on if we can more advanced techs that can generate more correct/complex patches, then they should largely refined!}}
\finding{\opad performs worse on developer patches, and should be applied with caution on (future) automated \apr tools that can generate more complicated patches.}
\Comment{\jun{Could we mention some potential causes, for example, it is known that test generation tool could not achieve high coverage. It's very possible that many tests even don't cover the revision of patches, so the opad technique will definitely not work in that case. If so, we can explain why only a very small subset of overfitting patches are detected.}}

%


\subsubsection{\patchsim{}}
Table~\ref{tab:patch-sim_final} presents the performance of \patchsim{} on previous datasets (i.e., \xiongdata{} from its original paper~\cite{xiong2018identifying} and \shangwendata{} from the previous study~\cite{DBLP:conf/kbse/WangWLWQZMJ20}) and our dataset. 

\begin{table}[h]
  \setlength\tabcolsep{6.0pt} 
  \def\arraystretch{1.0}
  \scriptsize
  \caption{Performance of  \patchsim{} on  different datasets}
  \centering
  \begin{tabular}{l|r|r|r|r|r|r|r}
\toprule
Datasets & \textit{TP} & \textit{FP} & \textit{TN} & \textit{FN} & \textit{\overfitprecision{}} & \textit{\overfitrecall{}} & \textit{\correctrecall} \\
\midrule
\xiongdata  & 62 & 0 & 29 & 48 & 100.00\% & 56.36\% & 100.00\% \\
\shangwendata & 249 & 51 & 186 & 392 & 83.00\% & 38.85\% & 78.48\% \\ \hline
\dfjoldprapr & 210 & 3 & 7 & 283 & 98.59\% & 42.60\% & 70.00\% \\
\bottomrule
\end{tabular}
  \label{tab:patch-sim_final}
\end{table}

\begin{listing}[h]
\begin{minted}{java}
-       if (fa * fb >= 0.0) {
+       if (fa * fb > 0.0D) {
          throw new ConvergenceException...}
\end{minted}
        
        \caption{A correct patch misclassified by \patchsim}
        \label{lst:patch-sim-fp-src}
\end{listing}

Note that due to an implementation issue~\cite{xiong2018identifying}, \patchsim{} does not support \dfjnew{} and subjects Closure and Mockito from \dfjold{}. 
Hence, following prior work~\cite{xiong2018identifying}, we only consider the patches supported by \patchsim{} from \dfjold{}. 
Overall, 493 overfitting and 10 correct patches in \dfjoldprapr{} are used. 
As shown in the table, \patchsim{} performs much worse on our dataset compared with initial \xiongdata{} results. 
For example, \patchsim{} achieves 100\% \overfitprecision{} and \correctrecall{} on its original dataset, whereas on \prapr{} patches, both metrics are degraded. 
Such a finding is similar to results in previous work~\cite{DBLP:conf/kbse/WangWLWQZMJ20}.
Listing~\ref{lst:patch-sim-fp-src} presents a sample FP that \patchsim{} mis-identifies the correct patch as overfitting. 
\patchsim{} is designed based on the \textit{mild consequence assumption} that a passing test should behave similarly on the correct patch and the buggy code. 
Therefore, the patch substantially changing the behavior of passing tests would be regarded as overfitting by \patchsim. 
In this example, the correct patch modifies the condition, leading to significant control-flow changes: on the buggy code, some passing tests can enter the \CodeIn{if} block and then throw the exception, whereas on the patch code these tests skip the \CodeIn{if} block and exception statement. 
Therefore, \patchsim{} considers the behaviors of these passing tests changed substantially, and further regards the patch as overfitting. 
However, correct patches are also likely to introduce large behavioral changes whereas overfitting patches can also induce tiny impact on passing tests.

%

\finding{The assumption that passing tests should behave similarly on correct patch code and buggy code can easily be broken when a larger patch space is considered, especially for the correct patches with non-trivial modifications on control flow.}

To further investigate the performance of \patchsim{}, 
Table~\ref{tab:dev_on_dynamics} presents the performance of \patchsim{} on 175
\Comment{\yuehan{Done here. We tested \patchsim on 214 patches for 4 categories(Chart, Lang, Math, Time) and 175 of them were tested successfully.}}
developer patches from \dfjold{} where \patchsim can be successfully applied. 
Interestingly, \patchsim{} performs even worse on developer patches with a \correctrecall{} of only 62.86\%, i.e., mis-classifying 37.14\% correct patches as incorrect. 
After manual inspection, we find that developer patches often involve sophisticated modifications and thus passing tests behave very differently from buggy code.
For example, Listing~\ref{lst:patch-sim-fp-dev} presents a developer patch which mis-identified as overfitting by \patchsim{}.
The patch modifies multiple lines, and several \CodeIn{if-else} statements. 
Thus, it significantly affects the paths
of some passing tests. 
In fact, it is prevalent that developer patches contain such sophisticated changes, and thus the assumption made by \patchsim{} can be easily violated on developer patches. 
Such results show that future \pcc{} work should exercise complex patches since advanced APR tools can generate more complex patches in the near future.
\begin{listing}[h]
\begin{minted}{java}
- if(... && index < seqEnd - 1 && ...) {
+ if(... && index < seqEnd - 2 && ...) {...}
+     if(start == seqEnd) {
+          return 0;
+     }
      ...
-    while(input.charAt(end) != ';') 
+    while(end<seqEnd && ((input.charAt(end)>='0' && input.charAt(end)<='9')
+        ||(input.charAt(end) >= 'a' && input.charAt(end) <= 'f')
+        ||(input.charAt(end) >= 'A' && input.charAt(end) <= 'F') ) )
         ...
+    boolean semiNext = (end != seqEnd) && (input.charAt(end) == ';');
-    return 2 + (end - start) + (isHex ? 1 : 0) + 1;
+    return 2 + (end - start) + (isHex ? 1 : 0) + (semiNext ? 1 : 0);
\end{minted}
        
        \caption{A developer patch misclassified by \patchsim}
        
        \label{lst:patch-sim-fp-dev}
\end{listing}

\finding{\Comment{\patchsim{} fails to identify almost all developers patches, since developers patches often involve complex modifications and \textit{mild consequence assumption} in \patchsim{} can easily be violated on these patches.} \patchsim{} misclassifies 30\% correct patches as incorrect on \dfjoldprapr{} and tends perform even worse with 38.14\% misclassifion rate on complicated developer patches. Our results also suggest that future dynamic \pcc{} techniques should consider more complex patches for evaluation.}

\section{Conclusion}

\Comment{Although various Patch-Correctness Checking (PCC) techniques have been proposed to address the important plausible patch problem, they are only evaluated on very limited datasets as the APR tools used for generating such patches can only explore a small subset of the possible patch search space, posing serious threats to external validity to existing PCC studies.} 
We have constructed a comprehensive PCC dataset for revisiting state-of-the-art PCC techniques. \Comment{Compared to prior PCC datasets which mainly leverage APR tools only capable of exploring a small subset of the possible patch search space,}\option{Our new dataset leverages the highly-optimized PraPR APR tool to return all possible plausible patches within its large predefined patch search space (including well-known fixing patterns from various prior APR tools) for more realistic/thorough PCC evaluation.}\Comment{ To our knowledge, this is the largest labeled dataset for PCC evaluation.}
Our study of PCC techniques on the new dataset has revealed various surprising findings and practical guidelines for future PCC work, \eg, various assumptions from prior studies~\cite{8918958,DBLP:conf/kbse/WangWLWQZMJ20,DBLP:conf/icse/WenCWHC18,DBLP:conf/kbse/XinR17/ssfix,le2017s3} no longer hold in our new dataset and semantics analysis should be included to determine patch correctness.
\bibliographystyle{acm}
\bibliography{main}
\end{document}